\pdfoutput=1
\documentclass[%
reprint,
superscriptaddress,
amsmath,amssymb,
aps,
showkeys
]{revtex4-2}

\usepackage{graphicx}
\usepackage{dcolumn}
\usepackage{bm}
\usepackage[utf8]{inputenc}
\usepackage[T1]{fontenc}
\usepackage{mathptmx}
\usepackage{svg}
\usepackage{amsmath}
\usepackage{caption}
\usepackage{subcaption}
\usepackage{xspace}

\usepackage{hyperref}
\hypersetup{colorlinks=true, linkcolor=blue, citecolor=blue, urlcolor=blue, unicode=false}
\newcommand{\overbar}[1]{\mkern 1.5mu\overline{\mkern-1.5mu#1\mkern-1.5mu}\mkern 1.5mu}

\begin{document}
	
	\preprint{APS/123-QED}
	\title{\textit{In-situ} exfoliation method of large-area 2D materials}

	\author{Antonija  Grubi\v{s}i\'{c}-\v{C}abo}
	\email{a.grubisic-cabo@rug.nl}
	\affiliation{ 
		Department of Applied Physics, KTH Royal Institute of Technology, Hannes Alfv\'{e}ns v\"{a}g 12, 114 19 Stockholm, Sweden
	}%
	\affiliation{Zernike Institute for Advanced Materials, University of Groningen, 9747 AG Groningen, The Netherlands}
	
	\author{Matteo Michiardi}
	\affiliation{Quantum Matter Institute, University of British Columbia, Vancouver, BC V6T 1Z4, Canada}
	\affiliation{Department of Physics and Astronomy, University of British Columbia, Vancouver, BC V6T 1Z1, Canada}
	
	\author{Charlotte E. Sanders}
	\affiliation{Central Laser Facility, STFC Rutherford Appleton Laboratory,
		Harwell 0X11 0QX, United Kingdom}
	
	\author{Marco Bianchi}
	\author{Davide Curcio}
	\affiliation{School of Physics and Astronomy, Aarhus University, 8000 C Aarhus, Denmark}
	
	\author{Dibya Phuyal}
	\author{Magnus H. Berntsen}
	\author{Qinda Guo}
	
	\author{Maciej Dendzik}
	\email{dendzik@kth.se}
	\affiliation{ 
		Department of Applied Physics, KTH Royal Institute of Technology, Hannes Alfv\'{e}ns v\"{a}g 12, 114 19 Stockholm, Sweden
	}%

	\date{\today}
	
	\begin{abstract}
	Two-dimensional (2D) materials provide an extremely rich platform to study novel physical phenomena arising from quantum confinement of charge carriers. Surface-sensitive techniques such as photoemission spectroscopy, scanning-tunnelling microscopy and electron diffraction, that work in ultra-high vacuum environment are prime techniques that have been employed with great success in unveiling new properties of 2D materials. The success in experimental studies of 2D materials, however, inherently relies on producing adsorbate-free, large-area, high-quality samples. The technique that most easily and readily yields 2D materials of highest quality is indubitably mechanical exfoliation from bulk-grown samples. However, as this technique is traditionally done in a dedicated environment, the transfer of samples into vacuum requires some form of surface cleaning that might tamper the samples' quality. In this article, we report on a simple and general method of \textit{in-situ} mechanical exfoliation directly in ultra-high vacuum that yields large-area single-layered films.  Multiple transition metal dichalcogenides, both metallic and semiconducting, are exfoliated \textit{in-situ} onto Au, Ag, and Ge. Exfoliated flakes are found to be sub-milimeter size with excellent crystallinity and purity, as evidenced by angle-resolved photoemission spectroscopy, atomic force microscopy and low-energy electron diffraction. In addition, we demonstrate exfoliation of air-sensitive 2D materials, surface alloys and possibility of controlling the substrate-2D material twist angle.

	\end{abstract}
	
	\maketitle
	
	
	\section{\label{sec:Introduciton}Introduction}
	The era of two-dimensional (2D) materials research has started with the discovery of graphene~\cite{Graphene_GeimNovoselov}-- a single layer of graphite that exhibits outstanding physical properties that do not exist in its bulk counterpart~\cite{Graphene_Kim_HallEff_BerryPhase, GeimGrapheneStatus,LimitsOfGrapheneDevices,GrapheneStrength,GrapheneChirality,GrapheneQuantumPhases}. The research that followed quickly expanded into other 2D materials beyond graphene, such as transition metal dichalcogenides (TMDCs)~\cite{2DTMDCsRevire_MBatzill,2DTMDCalloys,2DMoS2_KFMak2010,WS2_Au_SL_SingleCrystal_Luca2019, 2DTaS2_Charlotte,MoS2_Au_Jill,CVD_WSe2_RSC2018,WTe2_MBE3domains_ZX2017}, hexagonal boron nitride~\cite{2DhBN_Chhowalla,2DhBN_Sensor} and MX-enes~\cite{MXenesRise,MXenesForEnergyStorage}, to name a few. Among these, TMDCs are particularly interesting: single layer (SL) semiconducting TMDCs possess a direct band gap in the visible range~\cite{2DMoS2_KFMak2010,Splendiani10}, which makes them exceptional candidates for opto-electronic devices~\cite{TMDCspin_valleytronics, WS2StarkEffect_Gedik}, and metallic SL TMDCs showcase properties of 2D Mott physics, superconductivity and topological phases~\cite{Song22}. Furthermore, many physical properties of 2D materials are sensitive to careful interface engineering, through the stacking of layers into homo- and heterostructures as well as by adjusting the twist angle between adjacent layers~\cite{2DheterostructureReview, WS2TwistHomobilayer, TMDCinterfaceEngineering, MoS2Gr_Signe, MoS2GrHeterostruct}. 
    To effectively study and work with 2D materials, the research field must heavily rely on efficient methods of producing high-quality samples. The demanding requirements of size, that needs to be larger than the experimental probe, crystallinity and cleanliness of the interface, down to the atomic level, pose the most significant challenges. To maintain the materials in clean and optimal conditions \textit{in-situ} synthesis are usually favoured.
	
	One way to produce 2D materials is to employ well-established techniques such as chemical vapour deposition and molecular beam epitaxy to grow \textit{in-situ} ultra-thin films \cite{CVD_Review2021,CVD_MoS2_YHLee2012,WS2_Au_SL_SingleCrystal_Luca2019,CVD_MBE_V2S3_Au_Alex_2021,CVD_GrIr,CVD_WSe2_RSC2018,2Dmater_SynthesisChallenges2021}. The major drawbacks of epitaxial techniques are the complexity of defining new stable growth recipes, the choice of an appropriate substrate, and the sample mosaicity, which pose challenges that are unique for each material.
	On the other hand, mechanical exfoliation of 2D materials does not encounter the same obstacles, as the material is first grown in its bulk form and later exfoliated to the 2D limit, producing flakes of the highest quality \cite{ExfoliationProgress_InclAu2021, Exfoliation_AndCVD_Review2017}. Although conceptually simple, the exfoliation usually suffers from low yield of flakes small in size \cite{ExfoliationReview2021, ExfoliationProgress_InclAu2021}, stochastic distribution of the thickness, and residual contamination \cite{LayeredHeterostructure_Cleaning, Bubbles2_Smet2019}. Finding a solution to these problems is, thus, highly desirable for research in the field. Recently some pioneering work have shown that very large-area flakes of TMDCs can be deposited on polycrystalline gold substrates by exploiting the strong chemical affinity between the metal and the sample ~\cite{Au_2Dexfoliation_AdvMatter2016,Au_2D_MoS2_Exfol_ACSnano2018,Au_2Dexfoliation_Science2020, Au_assisted_2Dexfoliation_Nture2020,Au_ThermallyActivatedTransfer_Koch2020}.

	\begin{figure*}[tb]
		\centering
		\includegraphics[width=1\textwidth]{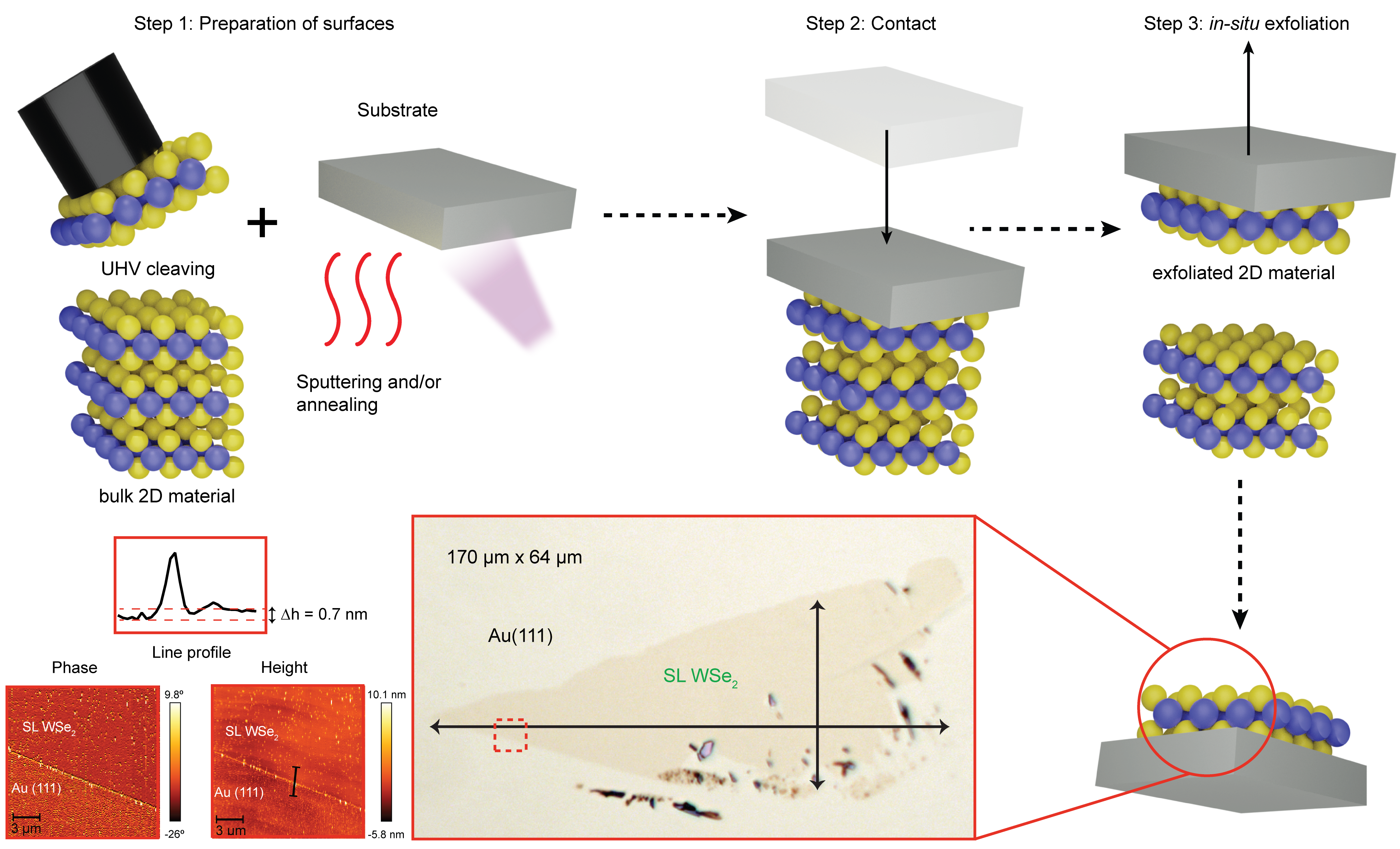}
		\caption{Sketch of the KISS exfoliation procedure. In the step 1, the sample surface is cleaved in UHV to expose an adsorbate-free surface, and the single crystal metal substrate is sputtered and annealed to generate an atomically-clean and flat surface. In step 2 the two surfaces are brought into contact. In step 3 sample and substrate are gently separated resulting in the \textit{in-situ} exfoliation of the 2D material onto the substrate. A high-quality, large-area 2D material is left on the substrate, as seen in the optical microscopy image for the case of single-layer (SL) WSe$_2$ on Au(111). The SL thickness of the sample is demonstrated by its characteristic height of 0.7~nm as measured by AFM data of the region marked with a dashed square in the optical image at the bottom}\label{Fig1:Sketch}
	\end{figure*}

	Here, we propose a new simple and generic method to efficiently exfoliate ultra-clean and large-area 2D materials: the kinetic \textit{in-situ} single-layer synthesis (KISS). The general idea of the KISS method is to bring into contact inside a vacuum environment atomically-clean surfaces of the bulk 2D material and the substrate establishing a chemical bonding that facilitates the exfoliation of single layers. We show that the KISS technique produces sub-millimeter flakes with excellent quality and purity onto both metallic and semiconducting substrates (Au(111), Ag(111), and Ge(100)). The substrate-2D material twist angle can be arbitrary chosen, which allows for exhaustive interface engineering of 2D-heterostructures. We employ Angle Resolved Photoemission Spectroscopy (ARPES), Low-energy Electron Diffraction (LEED), Atomic Force Microscopy (AFM) and optical microscopy to assess the sample quality and properties. This exfoliation procedure does not require specialised equipment beyond standard Ultra-High Vacuum (UHV) apparatus, and further surface-sensitive experiments are performed without any post-treatment.

	
	\section{\label{sec:Results}Results}
	\subsection{Description of the KISS method}
	Fig.~\ref{Fig1:Sketch} presents an overview of the KISS method. First, atomically-clean surfaces are prepared on both the substrate and the bulk 2D material using standard techniques directly in UHV; the substrate is prepared via cycles of ion sputtering and annealing, and the bulk 2D material is cleaved \textit{in-situ}. In the second step, the two surfaces are slowly brought into contact to establish a strong chemical bond between the 2D material and the substrate. In the last step the materials are slowly and rigidly separated resulting in the \textit{in-situ} exfoliation of the 2D material. Routinely obtained flakes of SL WSe$_2$ on Au(111) are sub-millimeter in size (see optical microscope image shown in Fig.~\ref{Fig1:Sketch}) and of excellent structural quality, as confirmed by LEED (Fig. S1 of the Supplementary Information). To determine the thickness of the flake we measure its height with AFM as shown in Fig.~\ref{Fig1:Sketch} and Fig. S2 of the Supplementary Information; the value of ca. 0.7~nm is in agreement with previously reported values for a SL WSe$_2$ \cite{MoS2_WSe2_Epitaxy_Science2015, WSe2_BL_sapphire_AFMheight_Li}.
	

	\subsection{Exfoliation of large-area SL WSe$_2$ on Ag(111)}
	
		\begin{figure*}[tbh!]
		\centering
		\includegraphics[width=1\textwidth]{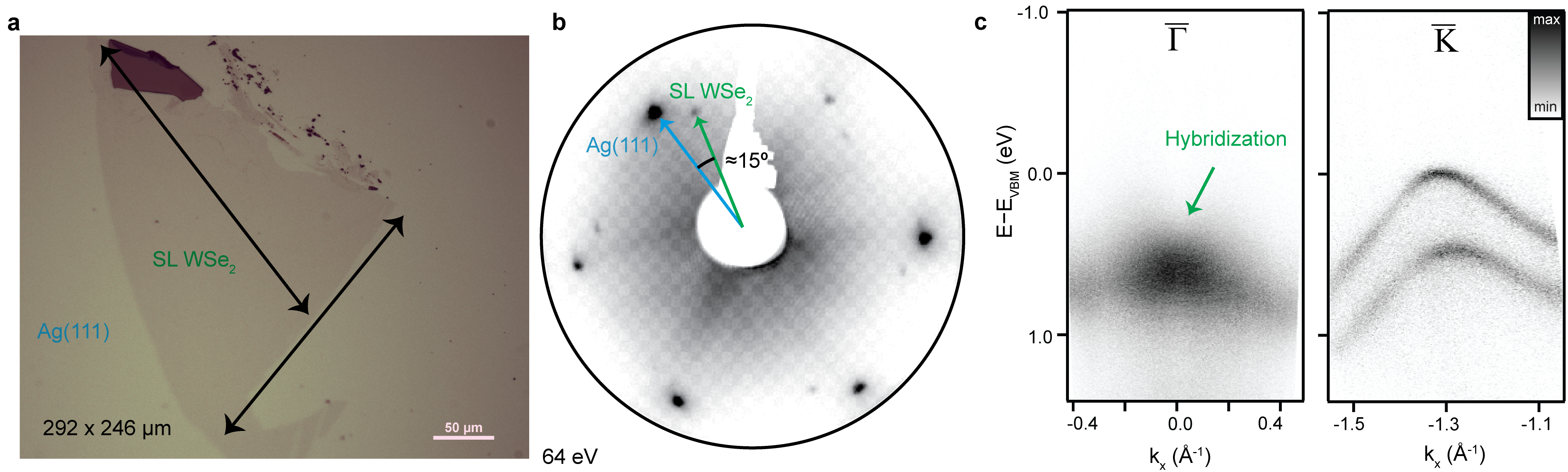}
		\caption{Silver-assisted exfoliation of a large-area SL WSe$_2$. \textbf{a} The optical microscopy image shows a flake of SL WSe$_2$ exfoliated by KISS method on Ag(111); the two major dimensions indicated by arrows are 292 and 246~$mu$m. Colour balance was adjusted to make the single layer more visible. The dark region on the top is to be associated to a small multi-layered region. \textbf{b} LEED image of the WSe$_2$ flake; the angle between the Bragg peaks of the sample and the substrate is ca. 15$^{\circ}$, confirming that the exfoliation angle can be arbitrarily chosen and it does not rely on coherent epitaxy. \textbf{c} ARPES data showing WSe$_2$ bands around $\overbar{\Gamma}$ (left) and $\overbar{\mathrm{K}}$ point (right). The energy is referenced to the valence band maximum (VBM) that is located at the $\overbar{\mathrm{K}}$ point. The green arrow indicates  where the hybridization between the WSe$_2$ and Ag(111) bands appear. The higher energy of the bands at the K point is characteristic of the single-layer nature of the flake}\label{Fig2:WSe2/Ag}
	\end{figure*}
	As the test case for the KISS method, SL WSe$_2$ was also exfoliated on Ag(111) and characterised in detail. WSe$_2$ crystal and Ag(111) substrate were prepared as described in section~\ref{sec:Methods}, and the exfoliation procedure is described above. The KISS-exfoliation resulted in a large-area SL WSe$_2$ flake of 292 $\mu$m $\times$ 246~$\mu$m,  Fig.~\ref{Fig2:WSe2/Ag}a. The SL is continuous and homogeneous, with smaller multilayer regions close to the edges (darker colour in the optical microscope image). LEED data confirm the single domain crystallinity of the exfoliated flake, as only a single set of the WSe$_2$ diffraction spots can be seen in Fig.~\ref{Fig2:WSe2/Ag}b. 
	The WSe2$_2$ flake is rotated by ca. 15$^{\circ}$ with respect to the underlying Ag(111) substrate, as indicated by the blue and green arrows.
	
We performed Angle-resolved photoemission spectroscopy (ARPES) measurements to further evaluate the sample thickness and quality (see Fig.~\ref{Fig2:WSe2/Ag}c). It must be noted that because the KISS technique is performed in UHV, the exfoliation happens \textit{in-situ}, and samples are transferred to the ARPES setup without being exposed to any gas or pollutants. Overall, the ARPES data around the $\overbar{\mathrm{K}}$ point is of exceptionally high quality for exfoliated flakes of 2D materials, ascertaining the high level of crystallinity, purity, and flatness. 
	As expected for the SL, the valence band maximum (VBM) is located at the $\overbar{\mathrm{K}}$ point, shown in Figure \ref{Fig2:WSe2/Ag}c(right)~\cite{WSe2_uARPES_ML_BL_bulk, WSe2_ThinFilmsMBE_SKMo}. The large Zeeman-like spin-slitting of approximately 466~meV at the $\overbar{\mathrm{K}}$ point is in good agreement with previously reported values \cite{WSe2_ThinFilmsMBE_SKMo}.
	A single valence band is found at the Brillouin zone center, $\overbar{\Gamma}$ point, shown in Fig.~\ref{Fig2:WSe2/Ag}c (left), confirming that the exfoliated material is indeed a SL~\cite{WSe2_uARPES_ML_BL_bulk,WSe2_ThinFilmsMBE_SKMo}. 
	One can learn a lot about the physical interaction between the substrate and the WSe$_2$ flake from a careful inspection of its electron dispersion: hybridization of the 2D material bands with the substrate indicates a strong chemical bond, which manifests as broader and gapped spectral features. The valence band of WSe$_2$ at $\overbar{\mathrm{K}}$ is mostly derived from in-plane orbitals and does not hybridize with the substrate, resulting in the sharp bands observed. Contrarily, the valence band at $\overbar{\Gamma}$ primarily derives from out-of-plane chalcogen $p_z$ and transition metal $d_{z^2}$ orbitals and is directly involved in the bond with the substrate ~\cite{Bruix16}. As evident in \ref{Fig2:WSe2/Ag}c, the band is significantly broader at $\overbar{\Gamma}$ where electron bands of Ag(111) are crossing causing the hybridization, which indicates that a strong chemical bond is established~\cite{WS2_Ag_metalTrans_Mac}. This TMDC-substrate interaction is stronger than the inter-layer interaction in bulk TMDCs, and it is responsible for the exfoliation of single-layers. 
	It is important to note that despite the chemisorption to the substrate, we did not observe preferential rotational alignments between TMDC flakes and the Ag(111) and Au(111) substrates; this is consistent with the radial symmetry of the out-of-plane orbitals participating in the bond. The natural implication is that the orientation of the flake can be arbitrarily chosen during the KISS, providing a way to chose "twist" angle at the interface.
	
	Overall, TMDC exfoliation on the surface of silver seems to yield more copious and larger flakes than on gold, likely because of the stronger sample/substrate interaction~\cite{WS2_Ag_metalTrans_Mac}. However, it is important to notice that the silver surface passivates more quickly in air, and a similar exfoliation process performed outside vacuum conditions might result impaired.

	\subsection{Survey of suitable substrates and TMDC materials}

		\begin{figure*}[tb!]
		\centering
		\includegraphics[width=1\textwidth]{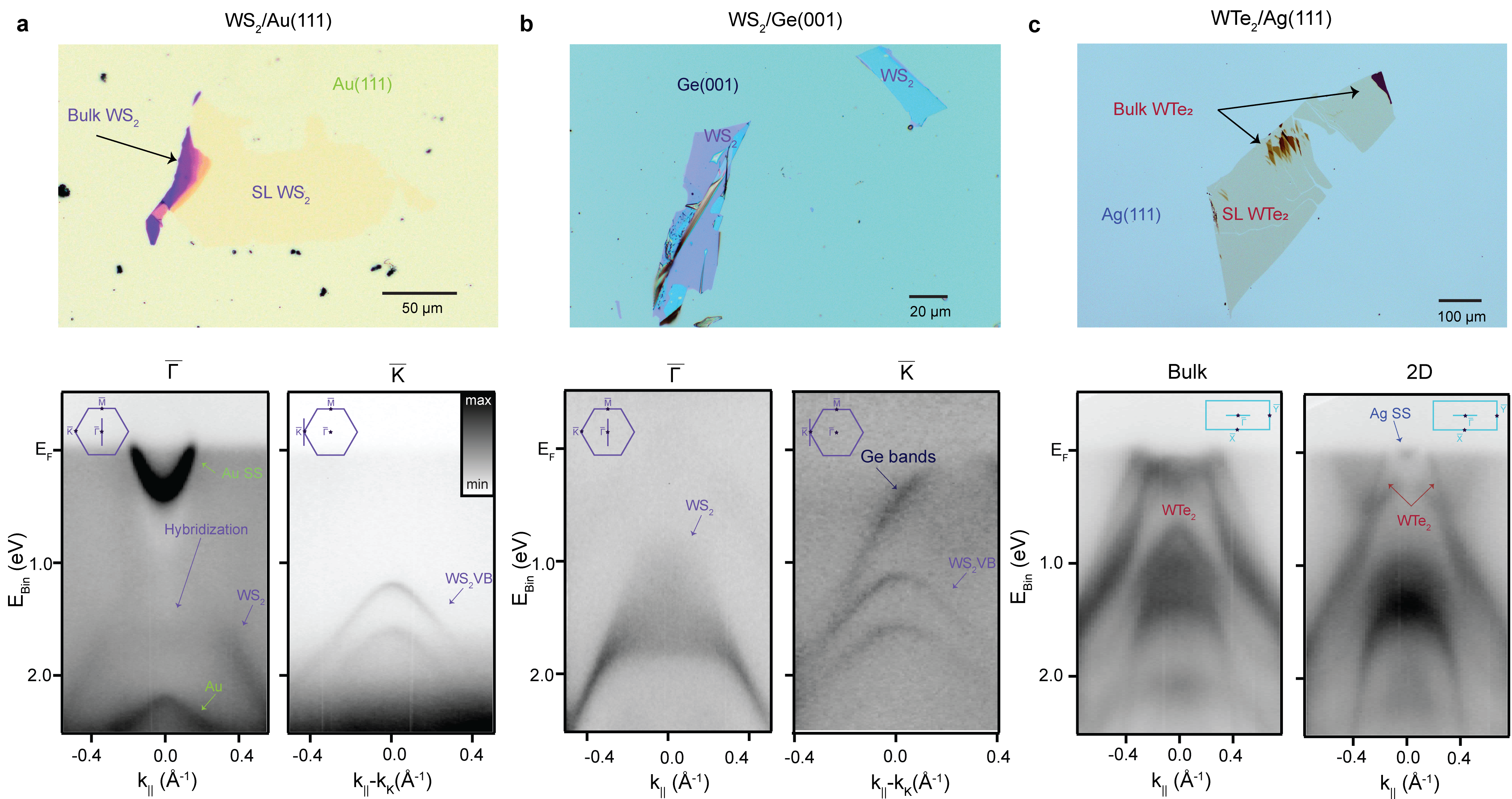}
		\caption{Universality of the KISS method: Exfoliation of different TMDC materials on multiple substrates. Optical microscopy images are shown on the top row and ARPES data acquired in-situ in the bottom row for three different systems. The insets illustrate the cuts in the momentum space that are plotted in the respective ARPES measurements. \textbf{a} Flakes of WS$_2$/Au(111) are over 100~$\mu$m in size and predominantly SL as demonstrated by the photoemission data where the valence band maximum is at the $\overbar{\mathrm{K}}$, and at $\overbar{\mathrm{\Gamma}}$ point the Au(111) surface state is visible. \textbf{b} Exfoliated flakes of WSe$_2$/Ge(001) demonstrate that exfoliation of 2D materials directly on semiconducting substrates is also possible. The quality of the ARPES data is uncompromised, however flakes are generally smaller and thicker than on metallic substrates. \textbf{c} 2D WTe$_2$ has been exfoliated on Ag(111) yielding flakes of hundreds of microns in size. The KISS exfoliation is particularly useful for air-sensitive samples, such as WTe$_2$, that can be exfoliated and measured \textit{in-situ}. The ARPES data is of remarkable quality and is shown on both a multilayered (3D) area of the flake (left) and on a 2D area (right). The spectra are markedly different especially close to the Fermi level highlighting the deep topological difference of the two cases.  }\label{Fig3:TMDCs/stuff}
	\end{figure*}

	We have explored the KISS method of exfoliation across various van-der-Waals layered materials and substrates. Fig.~\ref{Fig3:TMDCs/stuff} presents results for TMDCs WTe$_2$ and WS$2$, on substrates Au(111), Ag(111), and Ge(001). As shown in Fig.~\ref{Fig3:TMDCs/stuff}a the exfoliation of WS$_{2}$ yields similar results to the WSe$_{2}$ with a large SL flake and well-defined spectral features. Valence band spin-splitting of ca. 430~meV can be seen at $\overbar{\mathrm{K}}$, in agreement with the literature~\cite{WS2GrowthAu111_VBsplit, WS2Ag111_TRARPES_VBsplit}. The Au(111) surface state~\cite{111NobleMetalSS} is still clearly visible at the $\overbar{\Gamma}$ point; the strong signal indicates that the surface of the substrate is not damaged or contaminated with the KISS exfoliation. Additional measurements for MoS$_2$ can be found in the Supplementary Information and Fig. S3.
	
	Up to this point, we  have only considered noble metal substrates; however, semiconductors' surfaces prepared in UHV often host dangling bonds that can facilitate the adherence of the flake in a similar fashion that it does in metals. Insulating and semiconducting substrates are also preferred to access the pristine transport properties of 2D materials, and thus use of semiconducting substrates is highly desirable. Fig.~\ref{Fig3:TMDCs/stuff}b shows the results of the KISS-exfoliation of WS$_2$ on the (001) surface of the semiconductor germanium. Similar to the exfoliations using Ag(111) and Au(111), thin flakes of 2D WS$_2$ are observed on the surface. However, the average size of the flakes seems to be smaller, with a higher prevalence of multilayer regions.
 Altogether, these results support the breadth of applicability of the KISS method, with a broad potential pool of suitable semiconducting or metallic substrates with different surface orientations, that likely goes beyond what was tested in this work.

 	\begin{figure*}[tbh!]
		\centering
		\includegraphics[width=1\textwidth]{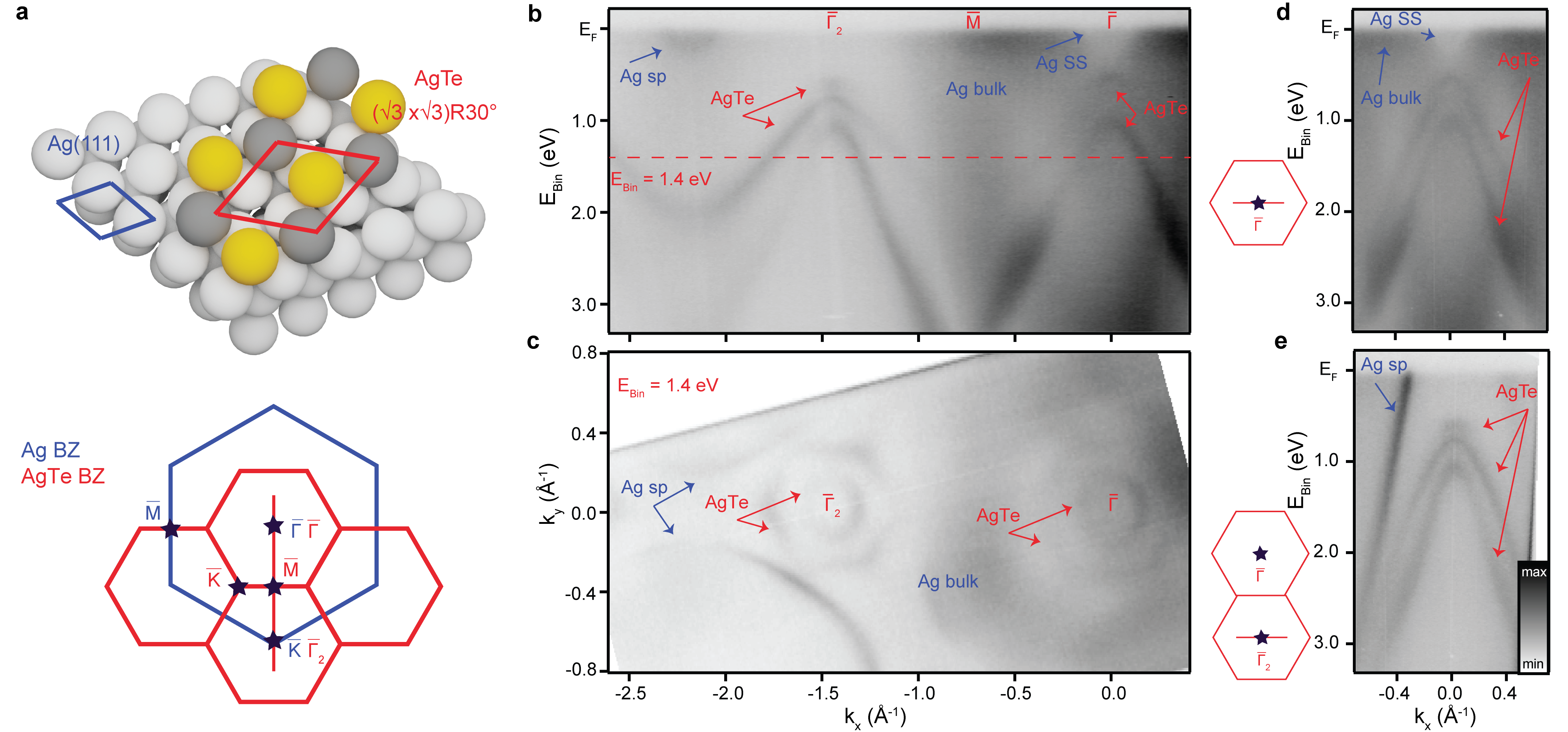}
		\caption{Formation of AgTe alloy on Ag(111). \textbf{a} Model of the crystal structure of the AgTe alloy on Ag(111) (top) and sketch of the corresponding surface Brillouin zones (bottom). \textbf{b} AgTe band structure along the $\overbar{\Gamma}$-$\overbar{\mathrm{K}}$ direction of the substrate. $\overbar{\Gamma}$, $\overbar{\mathrm{M}}$, $\overbar{\Gamma}_2$ mark the high-symmetry points of the AgTe. Dashed red line indicates where the constant energy contour in \textbf{c} is taken. \textbf{c} Constant energy contour taken at E$_{Bin}$=1.4~eV. Bulk and surface states (SS) of Ag(111) are visible, together with the AgTe bands. \textbf{d} and \textbf{e} show AgTe band structure taken at the $\overbar{\Gamma}$ point in the first surface Brillouin zone and $\overbar{\Gamma}_2$ point in the second surface Brillouin zone, respectively. The cuts' directions are indicated in the schematics next to the data. Blue and red arrows indicate Ag and AgTe bands, respectively.}
		\label{Fig4:AgTe_alloy}
	\end{figure*}
	
	Perhaps the most interesting applications of the KISS method is the exfoliation of metallic 2D materials. These systems exhibit fascinating electronic properties in their bulk form that have been subject to decades of studies, such as the superconductivity in NbSe$_2$ \cite{NbSe2_bulk_Supercond}, the remarkable charge density wave phases of TaSe$_2$ \cite{TaS2_CDW_1989,TaS2_CDW_Rossnagel,2DTaS2_Charlotte} or the Weyl semimetal state of WTe$_2$ \cite{WTe2_MBE3domains_ZX2017, WTe2_QSHE_realSpace}. In most cases, these materials cannot be exfoliated in air due to their debilitating air sensitivity, as their surface degrades outside of vacuum or inert atmosphere. This technical inconvenience poses a big challenge for studying their two-dimensional counterparts as the exfoliated material must be kept and transferred in an inert atmosphere at all times. The KISS exfoliation is optimal for these materials as it can be performed in UHV, directly in the experimental setup. We study the case of WTe$_2$ to prove its applicability. The exfoliation of WTe$_2$ flakes is generally afflicted not only by the air sensitivity but also by the low yield of small flakes, few-micrometer scale in size \cite{WTe2_exfoliation_ambient_unstable, WTe2_QSHE_realSpace,WTe2_Raman_MLtoBulk}. Fig.~\ref{Fig3:TMDCs/stuff}c presents the results of WTe$_2$ exfoliation on Ag(111), which produced very large flakes of predominantly SL thickness. ARPES data were acquired on regions with bulk (Fig.~\ref{Fig3:TMDCs/stuff}c (left)) and 2D (Fig.~\ref{Fig3:TMDCs/stuff}c (right)) character. In contrast to the previously reported 2D WTe$_2$ grown on bilayer graphene, where three rotational domains are present due to low substrate interaction and three-fold symmetry of the substrate~\cite{WTe2_MBE3domains_ZX2017}, only a single domain of WTe$_2$ is obtained using the KISS method (see Supplementary Fig. S4).
	We also notice that WTe$_2$ presented here appears to be $n$-doped, likely due to charge transfer from the underlying Ag(111). The flakes exfoliated by the KISS technique are hundreds of microns large, demonstrating that this approach allows for \textit{in-situ} study of large area flakes of metallic 2D materials.

	\subsection{2D materials beyond TMDCs}
	Unlike TMDCs or graphene, many 2D systems, such as silicene 
	\cite{Silicene_LeLay2012, Silicene_hybridized_NotDirac_Sanjoy2017}, 
	germanene \cite{ Germanene_LeLay_2014, Germanene_Ag111_2022}, bismuthene 
	\cite{Bismuthene_SiC_2017,Bismuthene_TRARPES_Mac22},
	or surface alloys, are stabilized by the interaction with the substrate. This presents significant difficulties in the synthesis of some of these materials. Previous successful attempts were realized exclusively via various growth methods, but here we demonstrate, that such materials can also be obtained using the proposed KISS technique.

	In particular, surface alloying is a recognised way to achieve unique physical and chemical properties not found in the bulk~\cite{SurfaceAlloys_Woodruf,SurfAlloys_Magnetic_Vasiliev_1997,2DTMDCalloys,AlloyFe_Au_onRu}. These include the appearance of Dirac nodal line fermions in CuSe grown on Cu(111)~\cite{Alloy_CuSe_onCu111}, a giant Rashba-type spin splitting in Ag$_2$Bi on Ag(111)~\cite{Alloy_Ag2Bi_Ag111_Ast,Alloy_Bi_Ag_Marco} and AgTe on Ag(111)~\cite{AgTE_Alloy, AgTe_AlloySTM2019} which shows presence of orbitally driven Rashba effect~\cite{AgTe_allyoRashbaSplit2020}. We speculate, that by controlling the temperature of a substrate during KISS exfoliation, one can prepare graphene analogues, such as bismuthene, germanine or silicene, or that one can perhaps realize novel phases of 2D materials. 
	
	Hexagonal 2D AgTe was synthesized by KISS exfoliation by approaching the Te rich surface of WTe$_2$ onto Ag(111) at a temperature of 373~K. The higher procedural temperature enables the mobile Te atoms to self-organise on the surface into the ($\sqrt{3}\times\sqrt{3}$)R30$^\circ$ superstructure depicted in Fig.~\ref{Fig4:AgTe_alloy}a. After the exfoliation we found regions containing amorphous Te clusters and large areas of high-quality AgTe on the substrate surface.  
	The symmetry of the AgTe and its electronic structure are in excellent agreement with results reported in Refs.~\cite{AgTE_Alloy,AgTe_AlloySTM2019,AgTe_allyoRashbaSplit2020} for samples grown with epitaxial techniques. The presence of bands related to the AgTe layer is evident in Fig.~\ref{Fig4:AgTe_alloy}b that shows a cut of the ARPES data along the $\overbar{\Gamma}$-$\overbar{\mathrm{K}}$ high symmetry direction of the Ag(111) lattice. The constant energy contour at a binding energy of 1.4~eV (Fig.~\ref{Fig4:AgTe_alloy}c) reveals Ag(111) bulk and surface states (blue arrows), and three-fold symmetric AgTe states (red arrows) around the $\overbar{\Gamma}$ and $\overbar{\mathrm{K}}$ points of the Ag(111) substrate surface Brillouin zone (SBZ). The observed dispersion of the AgTe states around these points is found to be identical (Fig.~\ref{Fig4:AgTe_alloy}d and e), which is consistent with the proposed ($\sqrt{3}\times\sqrt{3}$)R30$^\circ$ periodicity, as shown in the Fig.~\ref{Fig4:AgTe_alloy}a. The exact structure of the obtained AgTe is deduced by comparison of the measured electronic band structure with results of density functional theory calculations presented in~\cite{AgTE_Alloy}. High quality ARPES data and three-fold symmetry of the observed AgTe bands indicate that the layer is a single domain.
	
	These results bring to light an alternative for fast and straightforward synthesis of 2D materials in a manner that does not require either specialized equipment, such as source material evaporators, nor the time consuming outgassing and deposition processes necessary for molecular beam epitaxy growth.

	\section{\label{sec:Discussion}Summary}
In summary, we show a new method to produce large-area of atomically-flat flakes of 2D materials in UHV. This technique is versatile and applicable on a variety of TMDCs and different substrates. The flakes produced are single-domain and several hundreds of microns in size enabling direct \textit{in-situ} measurements with most spectroscopy, diffraction, microscopy and transport techniques. The in-plane angle between the flake and the substrate can be arbitrarily chosen enabling the \textit{in-situ} production of twisted heterostructures. Moreover, the possibility to exfoliate in a UHV environment allows for direct \textit{in-situ} study with ever-so-important techniques such as ARPES and scanning tunneling microscopy, even for air-sensitive samples that otherwise must be either grown in UHV or capped upon preparation in a glove box. We also showed the large-area exfoliation of metallic TMDCs that have so far remained elusive because of their air sensitivity. It is compelling to mention that the presented ARPES data were obtained using two experimental setups with macroscopic beam footprint, in contrast to typical ARPES measurements on exfoliated flakes that require nanofocusing capabilities, a feat available only in a handful of facilities in the world. Our results can provide a new platform to study a wide variety of new system that have so far been elusive, such as the topological superconductivity in twisted high-temperature superconductor heterostructures~\cite{Can21} or burgeoning field of 2D magnetism~\cite{Jiang21}.

	\section{\label{sec:Methods}Methods}

	\subsection*{Materials preparation}
	Au(111) single crystal (MaTecK), Au(111) thin film on mica ($\approx$200 nm thickness, Phasis) and Ag(111) thin film ($\approx$300 nm thickness, Georg-Albert-PVD) on mica were prepared by repeated cycles of Ar$^+$ (or Ne$^+$) sputtering (5 $\times$ 10$^{-5}$ mbar Ar, 1.7 kV, 30 min) and annealing (30 min at 600 K or 30 min at 800 K, depending on the setup). Ge(001) was prepared by Ar$^+$ (Ne$^+$) sputtering (3.4 $\times$ 10$^{-6}$ mbar Ar (Ne), 1.0 kV, 15 min) and annealing (15 min at 900 K).
	Bulk TMDC crystals (HQ graphene) were cleaved by the top-post method in the pressure better than 1 $\times$ 10$^{-8}$ mbar. Quality of the as cleaved vdW crystals was checked prior KISS-exfoliation by ARPES to confirm cleanliness and expected bulk structure. Bulk Au(111) was used for KISS-exfoliation shown in Fig. \ref{Fig1:Sketch}, while Au(111) and Ag(111) on mica were used for KISS-exfoliation shown in Figs. \ref{Fig2:WSe2/Ag}, \ref{Fig3:TMDCs/stuff} and \ref{Fig4:AgTe_alloy}.
	
	\subsection*{\textit{In-situ} Exfoliation (KISS method)}
	Samples with freshly cleaned surfaces were brought into direct contact in UHV by manual movement of a manipulator (micrometer screw) or transfer arm on which they were located. For KISS-exfoliation done at Baltazar facility, vdW material was kept on standard sample holder, while at the SGM3 beamline an appropriate spring loaded sample plate  was used for the vdW material. Following contact between the surfaces, substrate and a bulk crystal are brought out of contact by slow movement of the manipulator in the opposite direction. Exfoliated flakes were found on the substrate by the use of ARPES raster-mapping to locate signals originating from core levels or valence bands. See Supplementary Information and Supplementary Fig. S5. The KISS exfoliation was performed in pressures better than 5 $\times$ 10$^{-10}$ mbar. KISS-exfoliation was done at room temperature, except in the case of AgTe alloy where it was done at an elevated temperature of approximately 400 K.
	
	\subsection*{Low-Energy Electron Diffraction}
	Low-energy electron diffraction (LEED) measurements were performed following KISS exfoliation and taken at room temperature with an ErLEED 150, SPECS system, with spot size of ca. 1~mm. 
	
	\subsection*{Atomic Force Microscopy characterisation}
	AFM measurements were performed under ambient conditions with Dimension FastScan, Bruker and Cypher A AFM, Asylum Research  instruments. Measurements were performed using tapping mode, following ARPES and LEED measurements. AFM data was analysed using Gwyddion software ~\cite{Gwyddion}.
	
	\subsection*{Angle-resolved Photoemission Spectroscopy}
	ARPES measurements were performed at the BALTAZAR laboratory (KTH Royal Institute of Technology in Stockholm, Sweden)~\cite{BALTAZAR} and at the SGM3 beamline of the synchrotron radiation facility ASTRID2 (Aarhus University, Denmark)~\cite{SGM3beamline}. Measurements were performed at room temperature with base pressure better than 1 $\times$ 10$^{-10}$ mbar. At the BALTAZAR facility \cite{BALTAZAR}, a high-power femtosecond laser (Amplitude, Tangor 100) with an adjustable repetition rate (from a single shot to 40 MHz), providing infrared pulses centered at 1030~ nm, is used for high harmonic generation (HHG) in Ar gas. A repetition rate of 250~kHz, and the fifth harmonic of h$\nu$ = 18.1~eV, were used for the experiments. Measurements were performed with the ARTOF analyzer (SPECS, Themis 1000), with energy and angular resolution of 14~meV and 0.1$^{\circ}$, respectively. Beam footprint on the sample was approximately 100~$\mu$m. ARPES data obtained at the SGM3 beamline were acquired at photon energies of h$\nu$ = 49 eV, 53 eV and 63 eV, with energy and angular resolution better than 20~meV and 0.1$^{\circ}$, respectively. Spot size was 190 $\times$ 90~$\mu$m. All ARPES data was taken at room temperature. Data analysis was performed using Igor Pro (WaveMetrics, Lake Oswego, OR, USA) software.\\

	\section*{Data availability}
	Data available upon request.
	
	\begin{acknowledgments}
	 M.D. acknowledges financial support from the G{\"o}ran Gustafsson foundation. Q.G acknowledges the fellowship from Chinese scholarship council (No.201907930007). M. H. B. acknowledges support from the Magnus Bergvall Foundation. D.P. acknowledges the support of the Swedish Research Council under Grant No: 2020-00681. Part of the experiments was performed at the SGM3 endstation of ASTRID2 synchrotron, ISA, under project number ISA-22-1017. This research is funded in part by a QuantEmX grant from ICAM and the Gordon and Betty Moore Foundation through Grant GBMF9616 to MM. This work was supported by VILLUM FONDEN via the Centre of Excellence for Dirac Materials (Grant No. 11744) and the Independent Research Fund Denmark  (Grant No. 1026-00089B). We thank professor Youguo Shi (IoP,
CAS), for the high-quality WSe2 sample. We acknowledge insightful discussions with Andrea Damascelli, Philip Hofmann and Oscar Tjernberg.
	\end{acknowledgments}
	
	\section*{Author contributions}
	A.G.C., M.M., C.E.S., M.B., M.H.B, Q.G. and M.D performed ARPES measurements and A.G.C. and M.M. analyzed the data. A.G.C., D.P., Q.G. and M.D. performed AFM measurements. A.G.C., C.E.S., M.M., M.B. and M.D. performed optical imaging. M.B. and D.C. provided experimental support, designed and tested the holding device for the KISS method. A.G.C and M.D. wrote the manuscript with input and discussion from all co-authors. M.D. and A.G.C. conceived the project. M.D. provided necessary infrastructure and was responsible for overall project supervision. All authors discussed the results and their interpretation.
	\section*{Competing interests}
	
	The authors declare no competing interests.

	\bibliography{ref}

\begin{thebibliography}{78}%
\makeatletter
\providecommand \@ifxundefined [1]{%
 \@ifx{#1\undefined}
}%
\providecommand \@ifnum [1]{%
 \ifnum #1\expandafter \@firstoftwo
 \else \expandafter \@secondoftwo
 \fi
}%
\providecommand \@ifx [1]{%
 \ifx #1\expandafter \@firstoftwo
 \else \expandafter \@secondoftwo
 \fi
}%
\providecommand \natexlab [1]{#1}%
\providecommand \enquote  [1]{``#1''}%
\providecommand \bibnamefont  [1]{#1}%
\providecommand \bibfnamefont [1]{#1}%
\providecommand \citenamefont [1]{#1}%
\providecommand \href@noop [0]{\@secondoftwo}%
\providecommand \href [0]{\begingroup \@sanitize@url \@href}%
\providecommand \@href[1]{\@@startlink{#1}\@@href}%
\providecommand \@@href[1]{\endgroup#1\@@endlink}%
\providecommand \@sanitize@url [0]{\catcode `\\12\catcode `\$12\catcode
  `\&12\catcode `\#12\catcode `\^12\catcode `\_12\catcode `\%12\relax}%
\providecommand \@@startlink[1]{}%
\providecommand \@@endlink[0]{}%
\providecommand \url  [0]{\begingroup\@sanitize@url \@url }%
\providecommand \@url [1]{\endgroup\@href {#1}{\urlprefix }}%
\providecommand \urlprefix  [0]{URL }%
\providecommand \Eprint [0]{\href }%
\providecommand \doibase [0]{https://doi.org/}%
\providecommand \selectlanguage [0]{\@gobble}%
\providecommand \bibinfo  [0]{\@secondoftwo}%
\providecommand \bibfield  [0]{\@secondoftwo}%
\providecommand \translation [1]{[#1]}%
\providecommand \BibitemOpen [0]{}%
\providecommand \bibitemStop [0]{}%
\providecommand \bibitemNoStop [0]{.\EOS\space}%
\providecommand \EOS [0]{\spacefactor3000\relax}%
\providecommand \BibitemShut  [1]{\csname bibitem#1\endcsname}%
\let\auto@bib@innerbib\@empty
\bibitem [{\citenamefont {Novoselov}\ \emph {et~al.}(2004)\citenamefont
  {Novoselov}, \citenamefont {Geim}, \citenamefont {Morozov}, \citenamefont
  {Jiang}, \citenamefont {Zhang}, \citenamefont {Dubonos}, \citenamefont
  {Grigorieva},\ and\ \citenamefont {Firsov}}]{Graphene_GeimNovoselov}%
  \BibitemOpen
  \bibfield  {author} {\bibinfo {author} {\bibfnamefont {K.~S.}\ \bibnamefont
  {Novoselov}}, \bibinfo {author} {\bibfnamefont {A.~K.}\ \bibnamefont {Geim}},
  \bibinfo {author} {\bibfnamefont {S.~V.}\ \bibnamefont {Morozov}}, \bibinfo
  {author} {\bibfnamefont {D.}~\bibnamefont {Jiang}}, \bibinfo {author}
  {\bibfnamefont {Y.}~\bibnamefont {Zhang}}, \bibinfo {author} {\bibfnamefont
  {S.~V.}\ \bibnamefont {Dubonos}}, \bibinfo {author} {\bibfnamefont {I.~V.}\
  \bibnamefont {Grigorieva}},\ and\ \bibinfo {author} {\bibfnamefont {A.~A.}\
  \bibnamefont {Firsov}},\ }\bibfield  {title} {\bibinfo {title} {Electric
  field effect in atomically thin carbon films},\ }\href
  {https://doi.org/10.1126/science.1102896} {\bibfield  {journal} {\bibinfo
  {journal} {Science}\ }\textbf {\bibinfo {volume} {306}},\ \bibinfo {pages}
  {666} (\bibinfo {year} {2004})},\ \Eprint
  {https://arxiv.org/abs/https://www.science.org/doi/pdf/10.1126/science.1102896}
  {https://www.science.org/doi/pdf/10.1126/science.1102896} \BibitemShut
  {NoStop}%
\bibitem [{\citenamefont {Zhang}\ \emph {et~al.}(2005)\citenamefont {Zhang},
  \citenamefont {Tan}, \citenamefont {Stormer},\ and\ \citenamefont
  {Kim}}]{Graphene_Kim_HallEff_BerryPhase}%
  \BibitemOpen
  \bibfield  {author} {\bibinfo {author} {\bibfnamefont {Y.}~\bibnamefont
  {Zhang}}, \bibinfo {author} {\bibfnamefont {Y.-W.}\ \bibnamefont {Tan}},
  \bibinfo {author} {\bibfnamefont {H.~L.}\ \bibnamefont {Stormer}},\ and\
  \bibinfo {author} {\bibfnamefont {P.}~\bibnamefont {Kim}},\ }\bibfield
  {title} {\bibinfo {title} {Experimental observation of the quantum hall
  effect and {B}erry's phase in graphene},\ }\href
  {https://doi.org/10.1038/nature04235} {\bibfield  {journal} {\bibinfo
  {journal} {Nature}\ }\textbf {\bibinfo {volume} {438}},\ \bibinfo {pages}
  {201 } (\bibinfo {year} {2005})}\BibitemShut {NoStop}%
\bibitem [{\citenamefont {Geim}(2009)}]{GeimGrapheneStatus}%
  \BibitemOpen
  \bibfield  {author} {\bibinfo {author} {\bibfnamefont {A.~K.}\ \bibnamefont
  {Geim}},\ }\bibfield  {title} {\bibinfo {title} {Graphene: Status and
  prospects},\ }\href {https://doi.org/10.1126/science.1158877} {\bibfield
  {journal} {\bibinfo  {journal} {Science}\ }\textbf {\bibinfo {volume}
  {324}},\ \bibinfo {pages} {1530} (\bibinfo {year} {2009})},\ \Eprint
  {https://arxiv.org/abs/https://www.science.org/doi/pdf/10.1126/science.1158877}
  {https://www.science.org/doi/pdf/10.1126/science.1158877} \BibitemShut
  {NoStop}%
\bibitem [{\citenamefont {Chen}\ \emph {et~al.}(2008)\citenamefont {Chen},
  \citenamefont {Jang}, \citenamefont {Xiao}, \citenamefont {Ishigami},\ and\
  \citenamefont {Fuhrer}}]{LimitsOfGrapheneDevices}%
  \BibitemOpen
  \bibfield  {author} {\bibinfo {author} {\bibfnamefont {J.-H.}\ \bibnamefont
  {Chen}}, \bibinfo {author} {\bibfnamefont {C.}~\bibnamefont {Jang}}, \bibinfo
  {author} {\bibfnamefont {S.}~\bibnamefont {Xiao}}, \bibinfo {author}
  {\bibfnamefont {M.}~\bibnamefont {Ishigami}},\ and\ \bibinfo {author}
  {\bibfnamefont {M.~S.}\ \bibnamefont {Fuhrer}},\ }\bibfield  {title}
  {\bibinfo {title} {Intrinsic and extrinsic performance limits of graphene
  devices on {SiO}$_2$},\ }\href {https://doi.org/10.1038/nnano.2008.58}
  {\bibfield  {journal} {\bibinfo  {journal} {Nature Nanotechnology}\ }\textbf
  {\bibinfo {volume} {3}},\ \bibinfo {pages} {206 } (\bibinfo {year}
  {2008})}\BibitemShut {NoStop}%
\bibitem [{\citenamefont {Lee}\ \emph {et~al.}(2008)\citenamefont {Lee},
  \citenamefont {Wei}, \citenamefont {Kysar},\ and\ \citenamefont
  {Hone}}]{GrapheneStrength}%
  \BibitemOpen
  \bibfield  {author} {\bibinfo {author} {\bibfnamefont {C.}~\bibnamefont
  {Lee}}, \bibinfo {author} {\bibfnamefont {X.}~\bibnamefont {Wei}}, \bibinfo
  {author} {\bibfnamefont {J.~W.}\ \bibnamefont {Kysar}},\ and\ \bibinfo
  {author} {\bibfnamefont {J.}~\bibnamefont {Hone}},\ }\bibfield  {title}
  {\bibinfo {title} {Measurement of the elastic properties and intrinsic
  strength of monolayer graphene},\ }\href
  {https://doi.org/10.1126/science.1157996} {\bibfield  {journal} {\bibinfo
  {journal} {Science}\ }\textbf {\bibinfo {volume} {321}},\ \bibinfo {pages}
  {385} (\bibinfo {year} {2008})},\ \Eprint
  {https://arxiv.org/abs/https://www.science.org/doi/pdf/10.1126/science.1157996}
  {https://www.science.org/doi/pdf/10.1126/science.1157996} \BibitemShut
  {NoStop}%
\bibitem [{\citenamefont {Liu}\ \emph {et~al.}(2011)\citenamefont {Liu},
  \citenamefont {Bian}, \citenamefont {Miller},\ and\ \citenamefont
  {Chiang}}]{GrapheneChirality}%
  \BibitemOpen
  \bibfield  {author} {\bibinfo {author} {\bibfnamefont {Y.}~\bibnamefont
  {Liu}}, \bibinfo {author} {\bibfnamefont {G.}~\bibnamefont {Bian}}, \bibinfo
  {author} {\bibfnamefont {T.}~\bibnamefont {Miller}},\ and\ \bibinfo {author}
  {\bibfnamefont {T.-C.}\ \bibnamefont {Chiang}},\ }\bibfield  {title}
  {\bibinfo {title} {Visualizing electronic chirality and {B}erry phases in
  graphene systems using photoemission with circularly polarized light},\
  }\href {https://doi.org/10.1103/PhysRevLett.107.166803} {\bibfield  {journal}
  {\bibinfo  {journal} {Phys. Rev. Lett.}\ }\textbf {\bibinfo {volume} {107}},\
  \bibinfo {pages} {166803} (\bibinfo {year} {2011})}\BibitemShut {NoStop}%
\bibitem [{\citenamefont {Hwang}\ \emph {et~al.}(2011)\citenamefont {Hwang},
  \citenamefont {Park}, \citenamefont {Siegel}, \citenamefont {Fedorov},
  \citenamefont {Louie},\ and\ \citenamefont
  {Lanzara}}]{GrapheneQuantumPhases}%
  \BibitemOpen
  \bibfield  {author} {\bibinfo {author} {\bibfnamefont {C.}~\bibnamefont
  {Hwang}}, \bibinfo {author} {\bibfnamefont {C.-H.}\ \bibnamefont {Park}},
  \bibinfo {author} {\bibfnamefont {D.~A.}\ \bibnamefont {Siegel}}, \bibinfo
  {author} {\bibfnamefont {A.~V.}\ \bibnamefont {Fedorov}}, \bibinfo {author}
  {\bibfnamefont {S.~G.}\ \bibnamefont {Louie}},\ and\ \bibinfo {author}
  {\bibfnamefont {A.}~\bibnamefont {Lanzara}},\ }\bibfield  {title} {\bibinfo
  {title} {Direct measurement of quantum phases in graphene via photoemission
  spectroscopy},\ }\href {https://doi.org/10.1103/PhysRevB.84.125422}
  {\bibfield  {journal} {\bibinfo  {journal} {Phys. Rev. B}\ }\textbf {\bibinfo
  {volume} {84}},\ \bibinfo {pages} {125422} (\bibinfo {year}
  {2011})}\BibitemShut {NoStop}%
\bibitem [{\citenamefont {Lasek}\ \emph {et~al.}(2021)\citenamefont {Lasek},
  \citenamefont {Li}, \citenamefont {Kolekar}, \citenamefont {Coelho},
  \citenamefont {Guo}, \citenamefont {Zhang}, \citenamefont {Wang},\ and\
  \citenamefont {Batzill}}]{2DTMDCsRevire_MBatzill}%
  \BibitemOpen
  \bibfield  {author} {\bibinfo {author} {\bibfnamefont {K.}~\bibnamefont
  {Lasek}}, \bibinfo {author} {\bibfnamefont {J.}~\bibnamefont {Li}}, \bibinfo
  {author} {\bibfnamefont {S.}~\bibnamefont {Kolekar}}, \bibinfo {author}
  {\bibfnamefont {P.~M.}\ \bibnamefont {Coelho}}, \bibinfo {author}
  {\bibfnamefont {L.}~\bibnamefont {Guo}}, \bibinfo {author} {\bibfnamefont
  {M.}~\bibnamefont {Zhang}}, \bibinfo {author} {\bibfnamefont
  {Z.}~\bibnamefont {Wang}},\ and\ \bibinfo {author} {\bibfnamefont
  {M.}~\bibnamefont {Batzill}},\ }\bibfield  {title} {\bibinfo {title}
  {Synthesis and characterization of 2{D} transition metal dichalcogenides:
  Recent progress from a vacuum surface science perspective},\ }\href
  {https://doi.org/https://doi.org/10.1016/j.surfrep.2021.100523} {\bibfield
  {journal} {\bibinfo  {journal} {Surface Science Reports}\ }\textbf {\bibinfo
  {volume} {76}},\ \bibinfo {pages} {100523} (\bibinfo {year}
  {2021})}\BibitemShut {NoStop}%
\bibitem [{\citenamefont {Singh}\ \emph {et~al.}(2021)\citenamefont {Singh},
  \citenamefont {Kumbhakar}, \citenamefont {Krishnamoorthy}, \citenamefont
  {Nakano}, \citenamefont {Sadasivuni}, \citenamefont {Vashishta},
  \citenamefont {Roy}, \citenamefont {Kochat},\ and\ \citenamefont
  {Tiwary}}]{2DTMDCalloys}%
  \BibitemOpen
  \bibfield  {author} {\bibinfo {author} {\bibfnamefont {A.~K.}\ \bibnamefont
  {Singh}}, \bibinfo {author} {\bibfnamefont {P.}~\bibnamefont {Kumbhakar}},
  \bibinfo {author} {\bibfnamefont {A.}~\bibnamefont {Krishnamoorthy}},
  \bibinfo {author} {\bibfnamefont {A.}~\bibnamefont {Nakano}}, \bibinfo
  {author} {\bibfnamefont {K.~K.}\ \bibnamefont {Sadasivuni}}, \bibinfo
  {author} {\bibfnamefont {P.}~\bibnamefont {Vashishta}}, \bibinfo {author}
  {\bibfnamefont {A.~K.}\ \bibnamefont {Roy}}, \bibinfo {author} {\bibfnamefont
  {V.}~\bibnamefont {Kochat}},\ and\ \bibinfo {author} {\bibfnamefont {C.~S.}\
  \bibnamefont {Tiwary}},\ }\bibfield  {title} {\bibinfo {title} {Review of
  strategies toward the development of alloy two-dimensional (2{D}) transition
  metal dichalcogenides},\ }\href
  {https://doi.org/https://doi.org/10.1016/j.isci.2021.103532} {\bibfield
  {journal} {\bibinfo  {journal} {iScience}\ }\textbf {\bibinfo {volume}
  {24}},\ \bibinfo {pages} {103532} (\bibinfo {year} {2021})}\BibitemShut
  {NoStop}%
\bibitem [{\citenamefont {Mak}\ \emph {et~al.}(2010)\citenamefont {Mak},
  \citenamefont {Lee}, \citenamefont {Hone}, \citenamefont {Shan},\ and\
  \citenamefont {Heinz}}]{2DMoS2_KFMak2010}%
  \BibitemOpen
  \bibfield  {author} {\bibinfo {author} {\bibfnamefont {K.~F.}\ \bibnamefont
  {Mak}}, \bibinfo {author} {\bibfnamefont {C.}~\bibnamefont {Lee}}, \bibinfo
  {author} {\bibfnamefont {J.}~\bibnamefont {Hone}}, \bibinfo {author}
  {\bibfnamefont {J.}~\bibnamefont {Shan}},\ and\ \bibinfo {author}
  {\bibfnamefont {T.~F.}\ \bibnamefont {Heinz}},\ }\bibfield  {title} {\bibinfo
  {title} {Atomically thin {M}o{S}$_2$: A new direct-gap semiconductor},\
  }\href {https://doi.org/10.1103/PhysRevLett.105.136805} {\bibfield  {journal}
  {\bibinfo  {journal} {Phys. Rev. Lett.}\ }\textbf {\bibinfo {volume} {105}},\
  \bibinfo {pages} {136805} (\bibinfo {year} {2010})}\BibitemShut {NoStop}%
\bibitem [{\citenamefont {Bignardi}\ \emph {et~al.}(2019)\citenamefont
  {Bignardi}, \citenamefont {Lizzit}, \citenamefont {Bana}, \citenamefont
  {Travaglia}, \citenamefont {Lacovig}, \citenamefont {Sanders}, \citenamefont
  {Dendzik}, \citenamefont {Michiardi}, \citenamefont {Bianchi}, \citenamefont
  {Ewert}, \citenamefont {Bu\ss{}}, \citenamefont {Falta}, \citenamefont
  {Flege}, \citenamefont {Baraldi}, \citenamefont {Larciprete}, \citenamefont
  {Hofmann},\ and\ \citenamefont {Lizzit}}]{WS2_Au_SL_SingleCrystal_Luca2019}%
  \BibitemOpen
  \bibfield  {author} {\bibinfo {author} {\bibfnamefont {L.}~\bibnamefont
  {Bignardi}}, \bibinfo {author} {\bibfnamefont {D.}~\bibnamefont {Lizzit}},
  \bibinfo {author} {\bibfnamefont {H.}~\bibnamefont {Bana}}, \bibinfo {author}
  {\bibfnamefont {E.}~\bibnamefont {Travaglia}}, \bibinfo {author}
  {\bibfnamefont {P.}~\bibnamefont {Lacovig}}, \bibinfo {author} {\bibfnamefont
  {C.~E.}\ \bibnamefont {Sanders}}, \bibinfo {author} {\bibfnamefont
  {M.}~\bibnamefont {Dendzik}}, \bibinfo {author} {\bibfnamefont
  {M.}~\bibnamefont {Michiardi}}, \bibinfo {author} {\bibfnamefont
  {M.}~\bibnamefont {Bianchi}}, \bibinfo {author} {\bibfnamefont
  {M.}~\bibnamefont {Ewert}}, \bibinfo {author} {\bibfnamefont
  {L.}~\bibnamefont {Bu\ss{}}}, \bibinfo {author} {\bibfnamefont
  {J.}~\bibnamefont {Falta}}, \bibinfo {author} {\bibfnamefont {J.~I.}\
  \bibnamefont {Flege}}, \bibinfo {author} {\bibfnamefont {A.}~\bibnamefont
  {Baraldi}}, \bibinfo {author} {\bibfnamefont {R.}~\bibnamefont {Larciprete}},
  \bibinfo {author} {\bibfnamefont {P.}~\bibnamefont {Hofmann}},\ and\ \bibinfo
  {author} {\bibfnamefont {S.}~\bibnamefont {Lizzit}},\ }\bibfield  {title}
  {\bibinfo {title} {Growth and structure of singly oriented single-layer
  tungsten disulfide on {A}u(111)},\ }\href
  {https://doi.org/10.1103/PhysRevMaterials.3.014003} {\bibfield  {journal}
  {\bibinfo  {journal} {Phys. Rev. Materials}\ }\textbf {\bibinfo {volume}
  {3}},\ \bibinfo {pages} {014003} (\bibinfo {year} {2019})}\BibitemShut
  {NoStop}%
\bibitem [{\citenamefont {Sanders}\ \emph {et~al.}(2016)\citenamefont
  {Sanders}, \citenamefont {Dendzik}, \citenamefont {Ngankeu}, \citenamefont
  {Eich}, \citenamefont {Bruix}, \citenamefont {Bianchi}, \citenamefont {Miwa},
  \citenamefont {Hammer}, \citenamefont {Khajetoorians},\ and\ \citenamefont
  {Hofmann}}]{2DTaS2_Charlotte}%
  \BibitemOpen
  \bibfield  {author} {\bibinfo {author} {\bibfnamefont {C.~E.}\ \bibnamefont
  {Sanders}}, \bibinfo {author} {\bibfnamefont {M.}~\bibnamefont {Dendzik}},
  \bibinfo {author} {\bibfnamefont {A.~S.}\ \bibnamefont {Ngankeu}}, \bibinfo
  {author} {\bibfnamefont {A.}~\bibnamefont {Eich}}, \bibinfo {author}
  {\bibfnamefont {A.}~\bibnamefont {Bruix}}, \bibinfo {author} {\bibfnamefont
  {M.}~\bibnamefont {Bianchi}}, \bibinfo {author} {\bibfnamefont {J.~A.}\
  \bibnamefont {Miwa}}, \bibinfo {author} {\bibfnamefont {B.}~\bibnamefont
  {Hammer}}, \bibinfo {author} {\bibfnamefont {A.~A.}\ \bibnamefont
  {Khajetoorians}},\ and\ \bibinfo {author} {\bibfnamefont {P.}~\bibnamefont
  {Hofmann}},\ }\bibfield  {title} {\bibinfo {title} {Crystalline and
  electronic structure of single-layer {T}a{S}$_2$},\ }\href
  {https://doi.org/10.1103/PhysRevB.94.081404} {\bibfield  {journal} {\bibinfo
  {journal} {Phys. Rev. B}\ }\textbf {\bibinfo {volume} {94}},\ \bibinfo
  {pages} {081404} (\bibinfo {year} {2016})}\BibitemShut {NoStop}%
\bibitem [{\citenamefont {Miwa}\ \emph
  {et~al.}(2015{\natexlab{a}})\citenamefont {Miwa}, \citenamefont {Ulstrup},
  \citenamefont {S\o{}rensen}, \citenamefont {Dendzik}, \citenamefont
  {\ifmmode~\check{C}\else \v{C}\fi{}abo}, \citenamefont {Bianchi},
  \citenamefont {Lauritsen},\ and\ \citenamefont {Hofmann}}]{MoS2_Au_Jill}%
  \BibitemOpen
  \bibfield  {author} {\bibinfo {author} {\bibfnamefont {J.~A.}\ \bibnamefont
  {Miwa}}, \bibinfo {author} {\bibfnamefont {S.}~\bibnamefont {Ulstrup}},
  \bibinfo {author} {\bibfnamefont {S.~G.}\ \bibnamefont {S\o{}rensen}},
  \bibinfo {author} {\bibfnamefont {M.}~\bibnamefont {Dendzik}}, \bibinfo
  {author} {\bibfnamefont {A.~G. c. v. a.~c.}\ \bibnamefont
  {\ifmmode~\check{C}\else \v{C}\fi{}abo}}, \bibinfo {author} {\bibfnamefont
  {M.}~\bibnamefont {Bianchi}}, \bibinfo {author} {\bibfnamefont {J.~V.}\
  \bibnamefont {Lauritsen}},\ and\ \bibinfo {author} {\bibfnamefont
  {P.}~\bibnamefont {Hofmann}},\ }\bibfield  {title} {\bibinfo {title}
  {Electronic structure of epitaxial single-layer {M}o{S}$_2$},\ }\href
  {https://doi.org/10.1103/PhysRevLett.114.046802} {\bibfield  {journal}
  {\bibinfo  {journal} {Phys. Rev. Lett.}\ }\textbf {\bibinfo {volume} {114}},\
  \bibinfo {pages} {046802} (\bibinfo {year} {2015}{\natexlab{a}})}\BibitemShut
  {NoStop}%
\bibitem [{\citenamefont {Kang}\ \emph {et~al.}(2018)\citenamefont {Kang},
  \citenamefont {Lee}, \citenamefont {Yun}, \citenamefont {Song}, \citenamefont
  {Kim}, \citenamefont {Kim}, \citenamefont {Shin}, \citenamefont {Lee},
  \citenamefont {Heo}, \citenamefont {Kim}, \citenamefont {Lee},\ and\
  \citenamefont {Yu}}]{CVD_WSe2_RSC2018}%
  \BibitemOpen
  \bibfield  {author} {\bibinfo {author} {\bibfnamefont {W.~T.}\ \bibnamefont
  {Kang}}, \bibinfo {author} {\bibfnamefont {I.~M.}\ \bibnamefont {Lee}},
  \bibinfo {author} {\bibfnamefont {S.~J.}\ \bibnamefont {Yun}}, \bibinfo
  {author} {\bibfnamefont {Y.~I.}\ \bibnamefont {Song}}, \bibinfo {author}
  {\bibfnamefont {K.}~\bibnamefont {Kim}}, \bibinfo {author} {\bibfnamefont
  {D.-H.}\ \bibnamefont {Kim}}, \bibinfo {author} {\bibfnamefont {Y.~S.}\
  \bibnamefont {Shin}}, \bibinfo {author} {\bibfnamefont {K.}~\bibnamefont
  {Lee}}, \bibinfo {author} {\bibfnamefont {J.}~\bibnamefont {Heo}}, \bibinfo
  {author} {\bibfnamefont {Y.-M.}\ \bibnamefont {Kim}}, \bibinfo {author}
  {\bibfnamefont {Y.~H.}\ \bibnamefont {Lee}},\ and\ \bibinfo {author}
  {\bibfnamefont {W.~J.}\ \bibnamefont {Yu}},\ }\bibfield  {title} {\bibinfo
  {title} {Direct growth of doping controlled monolayer {WS}e2 by
  selenium-phosphorus substitution},\ }\href
  {https://doi.org/10.1039/C8NR03427C} {\bibfield  {journal} {\bibinfo
  {journal} {Nanoscale}\ }\textbf {\bibinfo {volume} {10}},\ \bibinfo {pages}
  {11397} (\bibinfo {year} {2018})}\BibitemShut {NoStop}%
\bibitem [{\citenamefont {Tang}\ \emph {et~al.}(2017)\citenamefont {Tang},
  \citenamefont {Zhang}, \citenamefont {Wong}, \citenamefont {Pedramrazi},
  \citenamefont {Tsai}, \citenamefont {Jia}, \citenamefont {Moritz},
  \citenamefont {Claassen}, \citenamefont {Ryu}, \citenamefont {Kahn},
  \citenamefont {Jiang}, \citenamefont {Yan}, \citenamefont {Hashimoto},
  \citenamefont {Lu}, \citenamefont {More}, \citenamefont {Hwang},
  \citenamefont {Hwang}, \citenamefont {Hussain}, \citenamefont {Chen},
  \citenamefont {Ugeda}, \citenamefont {Liu}, \citenamefont {Xie},
  \citenamefont {Devereaux}, \citenamefont {Crommie}, \citenamefont {Mo},\ and\
  \citenamefont {Shen}}]{WTe2_MBE3domains_ZX2017}%
  \BibitemOpen
  \bibfield  {author} {\bibinfo {author} {\bibfnamefont {S.}~\bibnamefont
  {Tang}}, \bibinfo {author} {\bibfnamefont {C.}~\bibnamefont {Zhang}},
  \bibinfo {author} {\bibfnamefont {D.}~\bibnamefont {Wong}}, \bibinfo {author}
  {\bibfnamefont {Z.}~\bibnamefont {Pedramrazi}}, \bibinfo {author}
  {\bibfnamefont {H.-Z.}\ \bibnamefont {Tsai}}, \bibinfo {author}
  {\bibfnamefont {C.}~\bibnamefont {Jia}}, \bibinfo {author} {\bibfnamefont
  {B.}~\bibnamefont {Moritz}}, \bibinfo {author} {\bibfnamefont
  {M.}~\bibnamefont {Claassen}}, \bibinfo {author} {\bibfnamefont
  {H.}~\bibnamefont {Ryu}}, \bibinfo {author} {\bibfnamefont {S.}~\bibnamefont
  {Kahn}}, \bibinfo {author} {\bibfnamefont {J.}~\bibnamefont {Jiang}},
  \bibinfo {author} {\bibfnamefont {H.}~\bibnamefont {Yan}}, \bibinfo {author}
  {\bibfnamefont {M.}~\bibnamefont {Hashimoto}}, \bibinfo {author}
  {\bibfnamefont {D.}~\bibnamefont {Lu}}, \bibinfo {author} {\bibfnamefont
  {R.~G.}\ \bibnamefont {More}}, \bibinfo {author} {\bibfnamefont {C.-C.}\
  \bibnamefont {Hwang}}, \bibinfo {author} {\bibfnamefont {C.}~\bibnamefont
  {Hwang}}, \bibinfo {author} {\bibfnamefont {Z.}~\bibnamefont {Hussain}},
  \bibinfo {author} {\bibfnamefont {Y.}~\bibnamefont {Chen}}, \bibinfo {author}
  {\bibfnamefont {M.~M.}\ \bibnamefont {Ugeda}}, \bibinfo {author}
  {\bibfnamefont {Z.}~\bibnamefont {Liu}}, \bibinfo {author} {\bibfnamefont
  {X.}~\bibnamefont {Xie}}, \bibinfo {author} {\bibfnamefont {T.~P.}\
  \bibnamefont {Devereaux}}, \bibinfo {author} {\bibfnamefont {M.~F.}\
  \bibnamefont {Crommie}}, \bibinfo {author} {\bibfnamefont {S.-K.}\
  \bibnamefont {Mo}},\ and\ \bibinfo {author} {\bibfnamefont {Z.-X.}\
  \bibnamefont {Shen}},\ }\bibfield  {title} {\bibinfo {title} {Quantum spin
  hall state in monolayer 1{T}'-{WT}e2},\ }\href
  {https://doi.org/10.1038/nphys4174} {\bibfield  {journal} {\bibinfo
  {journal} {Nature Physics}\ }\textbf {\bibinfo {volume} {13}},\ \bibinfo
  {pages} {683 } (\bibinfo {year} {2017})}\BibitemShut {NoStop}%
\bibitem [{\citenamefont {Ma}\ \emph {et~al.}(2022)\citenamefont {Ma},
  \citenamefont {Zhang}, \citenamefont {Jin}, \citenamefont {Wang},
  \citenamefont {Yoon}, \citenamefont {Hwang}, \citenamefont {Oh},
  \citenamefont {Jeong}, \citenamefont {Wang}, \citenamefont {Chatterjee},
  \citenamefont {Kim}, \citenamefont {Jang}, \citenamefont {Yang},
  \citenamefont {Ryu}, \citenamefont {Jeong}, \citenamefont {Ruoff},
  \citenamefont {Chhowalla}, \citenamefont {Ding},\ and\ \citenamefont
  {Shin}}]{2DhBN_Chhowalla}%
  \BibitemOpen
  \bibfield  {author} {\bibinfo {author} {\bibfnamefont {K.~Y.}\ \bibnamefont
  {Ma}}, \bibinfo {author} {\bibfnamefont {L.}~\bibnamefont {Zhang}}, \bibinfo
  {author} {\bibfnamefont {S.}~\bibnamefont {Jin}}, \bibinfo {author}
  {\bibfnamefont {Y.}~\bibnamefont {Wang}}, \bibinfo {author} {\bibfnamefont
  {S.~I.}\ \bibnamefont {Yoon}}, \bibinfo {author} {\bibfnamefont
  {H.}~\bibnamefont {Hwang}}, \bibinfo {author} {\bibfnamefont
  {J.}~\bibnamefont {Oh}}, \bibinfo {author} {\bibfnamefont {D.~S.}\
  \bibnamefont {Jeong}}, \bibinfo {author} {\bibfnamefont {M.}~\bibnamefont
  {Wang}}, \bibinfo {author} {\bibfnamefont {S.}~\bibnamefont {Chatterjee}},
  \bibinfo {author} {\bibfnamefont {G.}~\bibnamefont {Kim}}, \bibinfo {author}
  {\bibfnamefont {A.-R.}\ \bibnamefont {Jang}}, \bibinfo {author}
  {\bibfnamefont {J.}~\bibnamefont {Yang}}, \bibinfo {author} {\bibfnamefont
  {S.}~\bibnamefont {Ryu}}, \bibinfo {author} {\bibfnamefont {H.~Y.}\
  \bibnamefont {Jeong}}, \bibinfo {author} {\bibfnamefont {R.~S.}\ \bibnamefont
  {Ruoff}}, \bibinfo {author} {\bibfnamefont {M.}~\bibnamefont {Chhowalla}},
  \bibinfo {author} {\bibfnamefont {F.}~\bibnamefont {Ding}},\ and\ \bibinfo
  {author} {\bibfnamefont {H.~S.}\ \bibnamefont {Shin}},\ }\bibfield  {title}
  {\bibinfo {title} {Epitaxial single-crystal hexagonal boron nitride
  multilayers on ni (111)},\ }\href
  {https://doi.org/10.1038/s41586-022-04745-7} {\bibfield  {journal} {\bibinfo
  {journal} {Nature}\ }\textbf {\bibinfo {volume} {606}},\ \bibinfo {pages} {88
  } (\bibinfo {year} {2022})}\BibitemShut {NoStop}%
\bibitem [{\citenamefont {Khan}\ \emph {et~al.}(2016)\citenamefont {Khan},
  \citenamefont {Brownson}, \citenamefont {Randviir}, \citenamefont {Smith},\
  and\ \citenamefont {Banks}}]{2DhBN_Sensor}%
  \BibitemOpen
  \bibfield  {author} {\bibinfo {author} {\bibfnamefont {A.~F.}\ \bibnamefont
  {Khan}}, \bibinfo {author} {\bibfnamefont {D.~A.~C.}\ \bibnamefont
  {Brownson}}, \bibinfo {author} {\bibfnamefont {E.~P.}\ \bibnamefont
  {Randviir}}, \bibinfo {author} {\bibfnamefont {G.~C.}\ \bibnamefont
  {Smith}},\ and\ \bibinfo {author} {\bibfnamefont {C.~E.}\ \bibnamefont
  {Banks}},\ }\bibfield  {title} {\bibinfo {title} {2{D} hexagonal boron
  nitride (2{D}-h{BN}) explored for the electrochemical sensing of dopamine},\
  }\href {https://doi.org/10.1021/acs.analchem.6b02638} {\bibfield  {journal}
  {\bibinfo  {journal} {Analytical Chemistry}\ }\textbf {\bibinfo {volume}
  {88}},\ \bibinfo {pages} {9729} (\bibinfo {year} {2016})},\ \Eprint
  {https://arxiv.org/abs/https://doi.org/10.1021/acs.analchem.6b02638}
  {https://doi.org/10.1021/acs.analchem.6b02638} \BibitemShut {NoStop}%
\bibitem [{\citenamefont {Gogotsi}\ and\ \citenamefont
  {Anasori}(2019)}]{MXenesRise}%
  \BibitemOpen
  \bibfield  {author} {\bibinfo {author} {\bibfnamefont {Y.}~\bibnamefont
  {Gogotsi}}\ and\ \bibinfo {author} {\bibfnamefont {B.}~\bibnamefont
  {Anasori}},\ }\bibfield  {title} {\bibinfo {title} {The rise of {MX}enes},\
  }\href {https://doi.org/10.1021/acsnano.9b06394} {\bibfield  {journal}
  {\bibinfo  {journal} {ACS Nano}\ }\textbf {\bibinfo {volume} {13}},\ \bibinfo
  {pages} {8491} (\bibinfo {year} {2019})},\ \Eprint
  {https://arxiv.org/abs/https://doi.org/10.1021/acsnano.9b06394}
  {https://doi.org/10.1021/acsnano.9b06394} \BibitemShut {NoStop}%
\bibitem [{\citenamefont {Anasori}\ \emph {et~al.}(2017)\citenamefont
  {Anasori}, \citenamefont {Lukatskaya},\ and\ \citenamefont
  {Gogotsi}}]{MXenesForEnergyStorage}%
  \BibitemOpen
  \bibfield  {author} {\bibinfo {author} {\bibfnamefont {B.}~\bibnamefont
  {Anasori}}, \bibinfo {author} {\bibfnamefont {M.~R.}\ \bibnamefont
  {Lukatskaya}},\ and\ \bibinfo {author} {\bibfnamefont {Y.}~\bibnamefont
  {Gogotsi}},\ }\bibfield  {title} {\bibinfo {title} {2{D} metal carbides and
  nitrides ({MX}enes) for energy storage},\ }\href
  {https://doi.org/10.1038/natrevmats.2016.98} {\bibfield  {journal} {\bibinfo
  {journal} {Nature Reviews Materials}\ }\textbf {\bibinfo {volume} {2}},\
  \bibinfo {pages} {16098} (\bibinfo {year} {2017})}\BibitemShut {NoStop}%
\bibitem [{\citenamefont {Splendiani}\ \emph {et~al.}(2010)\citenamefont
  {Splendiani}, \citenamefont {Sun}, \citenamefont {Zhang}, \citenamefont {Li},
  \citenamefont {Kim}, \citenamefont {Chim}, \citenamefont {Galli},\ and\
  \citenamefont {Wang}}]{Splendiani10}%
  \BibitemOpen
  \bibfield  {author} {\bibinfo {author} {\bibfnamefont {A.}~\bibnamefont
  {Splendiani}}, \bibinfo {author} {\bibfnamefont {L.}~\bibnamefont {Sun}},
  \bibinfo {author} {\bibfnamefont {Y.}~\bibnamefont {Zhang}}, \bibinfo
  {author} {\bibfnamefont {T.}~\bibnamefont {Li}}, \bibinfo {author}
  {\bibfnamefont {J.}~\bibnamefont {Kim}}, \bibinfo {author} {\bibfnamefont
  {C.-Y.}\ \bibnamefont {Chim}}, \bibinfo {author} {\bibfnamefont
  {G.}~\bibnamefont {Galli}},\ and\ \bibinfo {author} {\bibfnamefont
  {F.}~\bibnamefont {Wang}},\ }\bibfield  {title} {\bibinfo {title} {Emerging
  photoluminescence in monolayer {M}o{S}$_2$},\ }\href
  {https://doi.org/10.1021/nl903868w} {\bibfield  {journal} {\bibinfo
  {journal} {Nano Letters}\ }\textbf {\bibinfo {volume} {10}},\ \bibinfo
  {pages} {1271} (\bibinfo {year} {2010})}\BibitemShut {NoStop}%
\bibitem [{\citenamefont {Zhong}\ \emph {et~al.}(2017)\citenamefont {Zhong},
  \citenamefont {Seyler}, \citenamefont {Linpeng}, \citenamefont {Cheng},
  \citenamefont {Sivadas}, \citenamefont {Huang}, \citenamefont {Schmidgall},
  \citenamefont {Taniguchi}, \citenamefont {Watanabe}, \citenamefont {McGuire},
  \citenamefont {Yao}, \citenamefont {Xiao}, \citenamefont {Fu},\ and\
  \citenamefont {Xu}}]{TMDCspin_valleytronics}%
  \BibitemOpen
  \bibfield  {author} {\bibinfo {author} {\bibfnamefont {D.}~\bibnamefont
  {Zhong}}, \bibinfo {author} {\bibfnamefont {K.~L.}\ \bibnamefont {Seyler}},
  \bibinfo {author} {\bibfnamefont {X.}~\bibnamefont {Linpeng}}, \bibinfo
  {author} {\bibfnamefont {R.}~\bibnamefont {Cheng}}, \bibinfo {author}
  {\bibfnamefont {N.}~\bibnamefont {Sivadas}}, \bibinfo {author} {\bibfnamefont
  {B.}~\bibnamefont {Huang}}, \bibinfo {author} {\bibfnamefont
  {E.}~\bibnamefont {Schmidgall}}, \bibinfo {author} {\bibfnamefont
  {T.}~\bibnamefont {Taniguchi}}, \bibinfo {author} {\bibfnamefont
  {K.}~\bibnamefont {Watanabe}}, \bibinfo {author} {\bibfnamefont {M.~A.}\
  \bibnamefont {McGuire}}, \bibinfo {author} {\bibfnamefont {W.}~\bibnamefont
  {Yao}}, \bibinfo {author} {\bibfnamefont {D.}~\bibnamefont {Xiao}}, \bibinfo
  {author} {\bibfnamefont {K.-M.~C.}\ \bibnamefont {Fu}},\ and\ \bibinfo
  {author} {\bibfnamefont {X.}~\bibnamefont {Xu}},\ }\bibfield  {title}
  {\bibinfo {title} {Van der waals engineering of ferromagnetic semiconductor
  heterostructures for spin and valleytronics},\ }\href
  {https://doi.org/10.1126/sciadv.1603113} {\bibfield  {journal} {\bibinfo
  {journal} {Science Advances}\ }\textbf {\bibinfo {volume} {3}},\ \bibinfo
  {pages} {e1603113} (\bibinfo {year} {2017})},\ \Eprint
  {https://arxiv.org/abs/https://www.science.org/doi/pdf/10.1126/sciadv.1603113}
  {https://www.science.org/doi/pdf/10.1126/sciadv.1603113} \BibitemShut
  {NoStop}%
\bibitem [{\citenamefont {Sie}\ \emph {et~al.}(2015)\citenamefont {Sie},
  \citenamefont {McIver}, \citenamefont {Lee}, \citenamefont {Fu},
  \citenamefont {Kong},\ and\ \citenamefont {Gedik}}]{WS2StarkEffect_Gedik}%
  \BibitemOpen
  \bibfield  {author} {\bibinfo {author} {\bibfnamefont {E.~J.}\ \bibnamefont
  {Sie}}, \bibinfo {author} {\bibfnamefont {J.~W.}\ \bibnamefont {McIver}},
  \bibinfo {author} {\bibfnamefont {Y.-H.}\ \bibnamefont {Lee}}, \bibinfo
  {author} {\bibfnamefont {L.}~\bibnamefont {Fu}}, \bibinfo {author}
  {\bibfnamefont {J.}~\bibnamefont {Kong}},\ and\ \bibinfo {author}
  {\bibfnamefont {N.}~\bibnamefont {Gedik}},\ }\bibfield  {title} {\bibinfo
  {title} {Valley-selective optical stark effect in monolayer {WS}$_2$},\
  }\href {https://doi.org/10.1038/nmat4156} {\bibfield  {journal} {\bibinfo
  {journal} {Nature Materials}\ }\textbf {\bibinfo {volume} {14}},\ \bibinfo
  {pages} {290 } (\bibinfo {year} {2015})}\BibitemShut {NoStop}%
\bibitem [{\citenamefont {Song}\ \emph {et~al.}(2022)\citenamefont {Song},
  \citenamefont {Li}, \citenamefont {Zhang},\ and\ \citenamefont
  {Shi}}]{Song22}%
  \BibitemOpen
  \bibfield  {author} {\bibinfo {author} {\bibfnamefont {L.}~\bibnamefont
  {Song}}, \bibinfo {author} {\bibfnamefont {H.}~\bibnamefont {Li}}, \bibinfo
  {author} {\bibfnamefont {Y.}~\bibnamefont {Zhang}},\ and\ \bibinfo {author}
  {\bibfnamefont {J.}~\bibnamefont {Shi}},\ }\bibfield  {title} {\bibinfo
  {title} {Recent progress of two-dimensional metallic transition metal
  dichalcogenides: Syntheses, physical properties, and applications},\ }\href
  {https://doi.org/10.1063/5.0083929} {\bibfield  {journal} {\bibinfo
  {journal} {Journal of Applied Physics}\ }\textbf {\bibinfo {volume} {131}},\
  \bibinfo {pages} {060902} (\bibinfo {year} {2022})}\BibitemShut {NoStop}%
\bibitem [{\citenamefont {Tan}\ \emph {et~al.}(2021)\citenamefont {Tan},
  \citenamefont {Li}, \citenamefont {Liu},\ and\ \citenamefont
  {Cheng}}]{2DheterostructureReview}%
  \BibitemOpen
  \bibfield  {author} {\bibinfo {author} {\bibfnamefont {J.}~\bibnamefont
  {Tan}}, \bibinfo {author} {\bibfnamefont {S.}~\bibnamefont {Li}}, \bibinfo
  {author} {\bibfnamefont {B.}~\bibnamefont {Liu}},\ and\ \bibinfo {author}
  {\bibfnamefont {H.-M.}\ \bibnamefont {Cheng}},\ }\bibfield  {title} {\bibinfo
  {title} {Structure, preparation, and applications of 2{D} material-based
  metal–semiconductor heterostructures},\ }\href
  {https://doi.org/https://doi.org/10.1002/sstr.202000093} {\bibfield
  {journal} {\bibinfo  {journal} {Small Structures}\ }\textbf {\bibinfo
  {volume} {2}},\ \bibinfo {pages} {2000093} (\bibinfo {year} {2021})},\
  \Eprint
  {https://arxiv.org/abs/https://onlinelibrary.wiley.com/doi/pdf/10.1002/sstr.202000093}
  {https://onlinelibrary.wiley.com/doi/pdf/10.1002/sstr.202000093} \BibitemShut
  {NoStop}%
\bibitem [{\citenamefont {Xu}\ \emph {et~al.}(2022)\citenamefont {Xu},
  \citenamefont {Duan}, \citenamefont {Liu}, \citenamefont {Wang},
  \citenamefont {Zhao}, \citenamefont {Li}, \citenamefont {Liu},\ and\
  \citenamefont {Liu}}]{WS2TwistHomobilayer}%
  \BibitemOpen
  \bibfield  {author} {\bibinfo {author} {\bibfnamefont {L.}~\bibnamefont
  {Xu}}, \bibinfo {author} {\bibfnamefont {W.}~\bibnamefont {Duan}}, \bibinfo
  {author} {\bibfnamefont {Y.}~\bibnamefont {Liu}}, \bibinfo {author}
  {\bibfnamefont {J.}~\bibnamefont {Wang}}, \bibinfo {author} {\bibfnamefont
  {Y.}~\bibnamefont {Zhao}}, \bibinfo {author} {\bibfnamefont {H.}~\bibnamefont
  {Li}}, \bibinfo {author} {\bibfnamefont {H.}~\bibnamefont {Liu}},\ and\
  \bibinfo {author} {\bibfnamefont {D.}~\bibnamefont {Liu}},\ }\bibfield
  {title} {\bibinfo {title} {Twist-angle-controlled neutral exciton
  annihilation in {WS}$_2$ homostructures},\ }\href
  {https://doi.org/10.1039/D2NR00195K} {\bibfield  {journal} {\bibinfo
  {journal} {Nanoscale}\ }\textbf {\bibinfo {volume} {14}},\ \bibinfo {pages}
  {5537} (\bibinfo {year} {2022})}\BibitemShut {NoStop}%
\bibitem [{\citenamefont {Li}\ and\ \citenamefont
  {Tang}(2020)}]{TMDCinterfaceEngineering}%
  \BibitemOpen
  \bibfield  {author} {\bibinfo {author} {\bibfnamefont {F.}~\bibnamefont
  {Li}}\ and\ \bibinfo {author} {\bibfnamefont {Q.}~\bibnamefont {Tang}},\
  }\bibfield  {title} {\bibinfo {title} {Modulating the electronic structure
  and in-plane activity of two-dimensional transition metal dichalcogenide
  ({M}o{S}2, {T}a{S}2, {N}b{S}2) monolayers by interfacial engineering},\
  }\href {https://doi.org/10.1021/acs.jpcc.0c01094} {\bibfield  {journal}
  {\bibinfo  {journal} {The Journal of Physical Chemistry C}\ }\textbf
  {\bibinfo {volume} {124}},\ \bibinfo {pages} {8822} (\bibinfo {year}
  {2020})},\ \Eprint
  {https://arxiv.org/abs/https://doi.org/10.1021/acs.jpcc.0c01094}
  {https://doi.org/10.1021/acs.jpcc.0c01094} \BibitemShut {NoStop}%
\bibitem [{\citenamefont {Miwa}\ \emph
  {et~al.}(2015{\natexlab{b}})\citenamefont {Miwa}, \citenamefont {Dendzik},
  \citenamefont {Grønborg}, \citenamefont {Bianchi}, \citenamefont
  {Lauritsen}, \citenamefont {Hofmann},\ and\ \citenamefont
  {Ulstrup}}]{MoS2Gr_Signe}%
  \BibitemOpen
  \bibfield  {author} {\bibinfo {author} {\bibfnamefont {J.~A.}\ \bibnamefont
  {Miwa}}, \bibinfo {author} {\bibfnamefont {M.}~\bibnamefont {Dendzik}},
  \bibinfo {author} {\bibfnamefont {S.~S.}\ \bibnamefont {Grønborg}}, \bibinfo
  {author} {\bibfnamefont {M.}~\bibnamefont {Bianchi}}, \bibinfo {author}
  {\bibfnamefont {J.~V.}\ \bibnamefont {Lauritsen}}, \bibinfo {author}
  {\bibfnamefont {P.}~\bibnamefont {Hofmann}},\ and\ \bibinfo {author}
  {\bibfnamefont {S.}~\bibnamefont {Ulstrup}},\ }\bibfield  {title} {\bibinfo
  {title} {Van der waals epitaxy of two-dimensional {M}o{S}$_2$–graphene
  heterostructures in ultrahigh vacuum},\ }\href
  {https://doi.org/10.1021/acsnano.5b02345} {\bibfield  {journal} {\bibinfo
  {journal} {ACS Nano}\ }\textbf {\bibinfo {volume} {9}},\ \bibinfo {pages}
  {6502} (\bibinfo {year} {2015}{\natexlab{b}})},\ \Eprint
  {https://arxiv.org/abs/https://doi.org/10.1021/acsnano.5b02345}
  {https://doi.org/10.1021/acsnano.5b02345} \BibitemShut {NoStop}%
\bibitem [{\citenamefont {Muñoz}\ \emph {et~al.}(2022)\citenamefont {Muñoz},
  \citenamefont {López-Elvira}, \citenamefont {Munuera}, \citenamefont
  {Frisenda}, \citenamefont {Sánchez-Sánchez}, \citenamefont {Ángel
  Martín-Gago},\ and\ \citenamefont
  {García-Hernández}}]{MoS2GrHeterostruct}%
  \BibitemOpen
  \bibfield  {author} {\bibinfo {author} {\bibfnamefont {R.}~\bibnamefont
  {Muñoz}}, \bibinfo {author} {\bibfnamefont {E.}~\bibnamefont
  {López-Elvira}}, \bibinfo {author} {\bibfnamefont {C.}~\bibnamefont
  {Munuera}}, \bibinfo {author} {\bibfnamefont {R.}~\bibnamefont {Frisenda}},
  \bibinfo {author} {\bibfnamefont {C.}~\bibnamefont {Sánchez-Sánchez}},
  \bibinfo {author} {\bibfnamefont {J.}~\bibnamefont {Ángel Martín-Gago}},\
  and\ \bibinfo {author} {\bibfnamefont {M.}~\bibnamefont
  {García-Hernández}},\ }\bibfield  {title} {\bibinfo {title} {Direct growth
  of graphene-{MoS}$_2$ heterostructure: Tailored interface for advanced
  devices},\ }\href
  {https://doi.org/https://doi.org/10.1016/j.apsusc.2021.151858} {\bibfield
  {journal} {\bibinfo  {journal} {Applied Surface Science}\ }\textbf {\bibinfo
  {volume} {581}},\ \bibinfo {pages} {151858} (\bibinfo {year}
  {2022})}\BibitemShut {NoStop}%
\bibitem [{\citenamefont {Sun}\ \emph {et~al.}(2021)\citenamefont {Sun},
  \citenamefont {Yuan}, \citenamefont {Gao}, \citenamefont {Yang},
  \citenamefont {Chhowalla}, \citenamefont {Gharahcheshmeh}, \citenamefont
  {Gleason}, \citenamefont {Choi}, \citenamefont {Hong},\ and\ \citenamefont
  {Liu}}]{CVD_Review2021}%
  \BibitemOpen
  \bibfield  {author} {\bibinfo {author} {\bibfnamefont {L.}~\bibnamefont
  {Sun}}, \bibinfo {author} {\bibfnamefont {G.}~\bibnamefont {Yuan}}, \bibinfo
  {author} {\bibfnamefont {L.}~\bibnamefont {Gao}}, \bibinfo {author}
  {\bibfnamefont {J.}~\bibnamefont {Yang}}, \bibinfo {author} {\bibfnamefont
  {M.}~\bibnamefont {Chhowalla}}, \bibinfo {author} {\bibfnamefont {M.~H.}\
  \bibnamefont {Gharahcheshmeh}}, \bibinfo {author} {\bibfnamefont {K.~K.}\
  \bibnamefont {Gleason}}, \bibinfo {author} {\bibfnamefont {Y.~S.}\
  \bibnamefont {Choi}}, \bibinfo {author} {\bibfnamefont {B.~H.}\ \bibnamefont
  {Hong}},\ and\ \bibinfo {author} {\bibfnamefont {Z.}~\bibnamefont {Liu}},\
  }\bibfield  {title} {\bibinfo {title} {Chemical vapour deposition},\ }\href
  {https://doi.org/10.1038/s43586-020-00005-y} {\bibfield  {journal} {\bibinfo
  {journal} {Nature Reviews Methods Primers}\ }\textbf {\bibinfo {volume}
  {1}},\ \bibinfo {pages} {5} (\bibinfo {year} {2021})}\BibitemShut {NoStop}%
\bibitem [{\citenamefont {Lee}\ \emph {et~al.}(2012)\citenamefont {Lee},
  \citenamefont {Zhang}, \citenamefont {Zhang}, \citenamefont {Chang},
  \citenamefont {Lin}, \citenamefont {Chang}, \citenamefont {Yu}, \citenamefont
  {Wang}, \citenamefont {Chang}, \citenamefont {Li},\ and\ \citenamefont
  {Lin}}]{CVD_MoS2_YHLee2012}%
  \BibitemOpen
  \bibfield  {author} {\bibinfo {author} {\bibfnamefont {Y.-H.}\ \bibnamefont
  {Lee}}, \bibinfo {author} {\bibfnamefont {X.-Q.}\ \bibnamefont {Zhang}},
  \bibinfo {author} {\bibfnamefont {W.}~\bibnamefont {Zhang}}, \bibinfo
  {author} {\bibfnamefont {M.-T.}\ \bibnamefont {Chang}}, \bibinfo {author}
  {\bibfnamefont {C.-T.}\ \bibnamefont {Lin}}, \bibinfo {author} {\bibfnamefont
  {K.-D.}\ \bibnamefont {Chang}}, \bibinfo {author} {\bibfnamefont {Y.-C.}\
  \bibnamefont {Yu}}, \bibinfo {author} {\bibfnamefont {J.~T.-W.}\ \bibnamefont
  {Wang}}, \bibinfo {author} {\bibfnamefont {C.-S.}\ \bibnamefont {Chang}},
  \bibinfo {author} {\bibfnamefont {L.-J.}\ \bibnamefont {Li}},\ and\ \bibinfo
  {author} {\bibfnamefont {T.-W.}\ \bibnamefont {Lin}},\ }\bibfield  {title}
  {\bibinfo {title} {Synthesis of large-area {M}o{S}$_2$ atomic layers with
  chemical vapor deposition},\ }\href
  {https://doi.org/https://doi.org/10.1002/adma.201104798} {\bibfield
  {journal} {\bibinfo  {journal} {Advanced Materials}\ }\textbf {\bibinfo
  {volume} {24}},\ \bibinfo {pages} {2320} (\bibinfo {year} {2012})},\ \Eprint
  {https://arxiv.org/abs/https://onlinelibrary.wiley.com/doi/pdf/10.1002/adm}
  {https://onlinelibrary.wiley.com/doi/pdf/10.1002/adm} \BibitemShut {NoStop}%
\bibitem [{\citenamefont {Kamber}\ \emph {et~al.}(2021)\citenamefont {Kamber},
  \citenamefont {Pakdel}, \citenamefont {Stan}, \citenamefont {Kamlapure},
  \citenamefont {Kiraly}, \citenamefont {Arnold}, \citenamefont {Eich},
  \citenamefont {Ngankeu}, \citenamefont {Bianchi}, \citenamefont {Miwa},
  \citenamefont {Sanders}, \citenamefont {Lanat\`a}, \citenamefont {Hofmann},\
  and\ \citenamefont {Khajetoorians}}]{CVD_MBE_V2S3_Au_Alex_2021}%
  \BibitemOpen
  \bibfield  {author} {\bibinfo {author} {\bibfnamefont {U.}~\bibnamefont
  {Kamber}}, \bibinfo {author} {\bibfnamefont {S.}~\bibnamefont {Pakdel}},
  \bibinfo {author} {\bibfnamefont {R.-M.}\ \bibnamefont {Stan}}, \bibinfo
  {author} {\bibfnamefont {A.}~\bibnamefont {Kamlapure}}, \bibinfo {author}
  {\bibfnamefont {B.}~\bibnamefont {Kiraly}}, \bibinfo {author} {\bibfnamefont
  {F.}~\bibnamefont {Arnold}}, \bibinfo {author} {\bibfnamefont
  {A.}~\bibnamefont {Eich}}, \bibinfo {author} {\bibfnamefont {A.~S.}\
  \bibnamefont {Ngankeu}}, \bibinfo {author} {\bibfnamefont {M.}~\bibnamefont
  {Bianchi}}, \bibinfo {author} {\bibfnamefont {J.~A.}\ \bibnamefont {Miwa}},
  \bibinfo {author} {\bibfnamefont {C.~E.}\ \bibnamefont {Sanders}}, \bibinfo
  {author} {\bibfnamefont {N.}~\bibnamefont {Lanat\`a}}, \bibinfo {author}
  {\bibfnamefont {P.}~\bibnamefont {Hofmann}},\ and\ \bibinfo {author}
  {\bibfnamefont {A.~A.}\ \bibnamefont {Khajetoorians}},\ }\bibfield  {title}
  {\bibinfo {title} {Moir\'e-induced electronic structure modifications in
  monolayer {V}$_2${S}$_3$ on {A}u(111)},\ }\href
  {https://doi.org/10.1103/PhysRevB.103.115414} {\bibfield  {journal} {\bibinfo
   {journal} {Phys. Rev. B}\ }\textbf {\bibinfo {volume} {103}},\ \bibinfo
  {pages} {115414} (\bibinfo {year} {2021})}\BibitemShut {NoStop}%
\bibitem [{\citenamefont {Coraux}\ \emph {et~al.}(2009)\citenamefont {Coraux},
  \citenamefont {T~N'Diaye}, \citenamefont {Engler}, \citenamefont {Busse},
  \citenamefont {Wall}, \citenamefont {Buckanie}, \citenamefont {Meyer~zu
  Heringdorf}, \citenamefont {van Gastel}, \citenamefont {Poelsema},\ and\
  \citenamefont {Michely}}]{CVD_GrIr}%
  \BibitemOpen
  \bibfield  {author} {\bibinfo {author} {\bibfnamefont {J.}~\bibnamefont
  {Coraux}}, \bibinfo {author} {\bibfnamefont {A.}~\bibnamefont {T~N'Diaye}},
  \bibinfo {author} {\bibfnamefont {M.}~\bibnamefont {Engler}}, \bibinfo
  {author} {\bibfnamefont {C.}~\bibnamefont {Busse}}, \bibinfo {author}
  {\bibfnamefont {D.}~\bibnamefont {Wall}}, \bibinfo {author} {\bibfnamefont
  {N.}~\bibnamefont {Buckanie}}, \bibinfo {author} {\bibfnamefont {F.-J.}\
  \bibnamefont {Meyer~zu Heringdorf}}, \bibinfo {author} {\bibfnamefont
  {R.}~\bibnamefont {van Gastel}}, \bibinfo {author} {\bibfnamefont
  {B.}~\bibnamefont {Poelsema}},\ and\ \bibinfo {author} {\bibfnamefont
  {T.}~\bibnamefont {Michely}},\ }\bibfield  {title} {\bibinfo {title} {Growth
  of graphene on ir(111)},\ }\href
  {https://doi.org/10.1088/1367-2630/11/2/023006} {\bibfield  {journal}
  {\bibinfo  {journal} {New Journal of Physics}\ }\textbf {\bibinfo {volume}
  {11}},\ \bibinfo {pages} {023006} (\bibinfo {year} {2009})}\BibitemShut
  {NoStop}%
\bibitem [{\citenamefont {Zhang}\ \emph
  {et~al.}(2021{\natexlab{a}})\citenamefont {Zhang}, \citenamefont {Dong},\
  and\ \citenamefont {Ding}}]{2Dmater_SynthesisChallenges2021}%
  \BibitemOpen
  \bibfield  {author} {\bibinfo {author} {\bibfnamefont {L.}~\bibnamefont
  {Zhang}}, \bibinfo {author} {\bibfnamefont {J.}~\bibnamefont {Dong}},\ and\
  \bibinfo {author} {\bibfnamefont {F.}~\bibnamefont {Ding}},\ }\bibfield
  {title} {\bibinfo {title} {Strategies, status, and challenges in wafer scale
  single crystalline two-dimensional materials synthesis},\ }\href
  {https://doi.org/10.1021/acs.chemrev.0c01191} {\bibfield  {journal} {\bibinfo
   {journal} {Chemical Reviews}\ }\textbf {\bibinfo {volume} {121}},\ \bibinfo
  {pages} {6321} (\bibinfo {year} {2021}{\natexlab{a}})},\ \Eprint
  {https://arxiv.org/abs/https://doi.org/10.1021/acs.chemrev.0c01191}
  {https://doi.org/10.1021/acs.chemrev.0c01191} \BibitemShut {NoStop}%
\bibitem [{\citenamefont {Liu}(2021)}]{ExfoliationProgress_InclAu2021}%
  \BibitemOpen
  \bibfield  {author} {\bibinfo {author} {\bibfnamefont {F.}~\bibnamefont
  {Liu}},\ }\bibfield  {title} {\bibinfo {title} {Mechanical exfoliation of
  large area 2{D} materials from vd{W} crystals},\ }\href
  {https://doi.org/https://doi.org/10.1016/j.progsurf.2021.100626} {\bibfield
  {journal} {\bibinfo  {journal} {Progress in Surface Science}\ }\textbf
  {\bibinfo {volume} {96}},\ \bibinfo {pages} {100626} (\bibinfo {year}
  {2021})}\BibitemShut {NoStop}%
\bibitem [{\citenamefont {Dong}\ and\ \citenamefont
  {Kuljanishvili}(2017)}]{Exfoliation_AndCVD_Review2017}%
  \BibitemOpen
  \bibfield  {author} {\bibinfo {author} {\bibfnamefont {R.}~\bibnamefont
  {Dong}}\ and\ \bibinfo {author} {\bibfnamefont {I.}~\bibnamefont
  {Kuljanishvili}},\ }\bibfield  {title} {\bibinfo {title} {Review article:
  Progress in fabrication of transition metal dichalcogenides heterostructure
  systems},\ }\href {https://doi.org/10.1116/1.4982736} {\bibfield  {journal}
  {\bibinfo  {journal} {Journal of Vacuum Science \& Technology B}\ }\textbf
  {\bibinfo {volume} {35}},\ \bibinfo {pages} {030803} (\bibinfo {year}
  {2017})},\ \Eprint {https://arxiv.org/abs/https://doi.org/10.1116/1.4982736}
  {https://doi.org/10.1116/1.4982736} \BibitemShut {NoStop}%
\bibitem [{\citenamefont {Zhang}\ \emph
  {et~al.}(2021{\natexlab{b}})\citenamefont {Zhang}, \citenamefont {Li},
  \citenamefont {Mu}, \citenamefont {Bai}, \citenamefont {Sun}, \citenamefont
  {Zhao}, \citenamefont {Zhang}, \citenamefont {Shan},\ and\ \citenamefont
  {Yang}}]{ExfoliationReview2021}%
  \BibitemOpen
  \bibfield  {author} {\bibinfo {author} {\bibfnamefont {X.}~\bibnamefont
  {Zhang}}, \bibinfo {author} {\bibfnamefont {Y.}~\bibnamefont {Li}}, \bibinfo
  {author} {\bibfnamefont {W.}~\bibnamefont {Mu}}, \bibinfo {author}
  {\bibfnamefont {W.}~\bibnamefont {Bai}}, \bibinfo {author} {\bibfnamefont
  {X.}~\bibnamefont {Sun}}, \bibinfo {author} {\bibfnamefont {M.}~\bibnamefont
  {Zhao}}, \bibinfo {author} {\bibfnamefont {Z.}~\bibnamefont {Zhang}},
  \bibinfo {author} {\bibfnamefont {F.}~\bibnamefont {Shan}},\ and\ \bibinfo
  {author} {\bibfnamefont {Z.}~\bibnamefont {Yang}},\ }\bibfield  {title}
  {\bibinfo {title} {Advanced tape-exfoliated method for preparing large-area
  {2D} monolayers: a review},\ }\href
  {https://doi.org/10.1088/2053-1583/ac016f} {\bibfield  {journal} {\bibinfo
  {journal} {2D Materials}\ }\textbf {\bibinfo {volume} {8}},\ \bibinfo {pages}
  {032002} (\bibinfo {year} {2021}{\natexlab{b}})}\BibitemShut {NoStop}%
\bibitem [{\citenamefont {Purdie}\ \emph {et~al.}(2018)\citenamefont {Purdie},
  \citenamefont {Pugno}, \citenamefont {Taniguchi}, \citenamefont {Watanabe},
  \citenamefont {Ferrari},\ and\ \citenamefont
  {Lombardo}}]{LayeredHeterostructure_Cleaning}%
  \BibitemOpen
  \bibfield  {author} {\bibinfo {author} {\bibfnamefont {D.~G.}\ \bibnamefont
  {Purdie}}, \bibinfo {author} {\bibfnamefont {N.~M.}\ \bibnamefont {Pugno}},
  \bibinfo {author} {\bibfnamefont {T.}~\bibnamefont {Taniguchi}}, \bibinfo
  {author} {\bibfnamefont {K.}~\bibnamefont {Watanabe}}, \bibinfo {author}
  {\bibfnamefont {A.~C.}\ \bibnamefont {Ferrari}},\ and\ \bibinfo {author}
  {\bibfnamefont {A.}~\bibnamefont {Lombardo}},\ }\bibfield  {title} {\bibinfo
  {title} {Cleaning interfaces in layered materials heterostructures},\ }\href
  {https://doi.org/10.1038/s41467-018-07558-3} {\bibfield  {journal} {\bibinfo
  {journal} {Nature Communications}\ }\textbf {\bibinfo {volume} {9}},\
  \bibinfo {pages} {5387} (\bibinfo {year} {2018})}\BibitemShut {NoStop}%
\bibitem [{\citenamefont {Kim}\ \emph {et~al.}(2019)\citenamefont {Kim},
  \citenamefont {Herlinger}, \citenamefont {Taniguchi}, \citenamefont
  {Watanabe},\ and\ \citenamefont {Smet}}]{Bubbles2_Smet2019}%
  \BibitemOpen
  \bibfield  {author} {\bibinfo {author} {\bibfnamefont {Y.}~\bibnamefont
  {Kim}}, \bibinfo {author} {\bibfnamefont {P.}~\bibnamefont {Herlinger}},
  \bibinfo {author} {\bibfnamefont {T.}~\bibnamefont {Taniguchi}}, \bibinfo
  {author} {\bibfnamefont {K.}~\bibnamefont {Watanabe}},\ and\ \bibinfo
  {author} {\bibfnamefont {J.~H.}\ \bibnamefont {Smet}},\ }\bibfield  {title}
  {\bibinfo {title} {Reliable postprocessing improvement of van der waals
  heterostructures},\ }\href {https://doi.org/10.1021/acsnano.9b06992}
  {\bibfield  {journal} {\bibinfo  {journal} {ACS Nano}\ }\textbf {\bibinfo
  {volume} {13}},\ \bibinfo {pages} {14182} (\bibinfo {year} {2019})},\ \Eprint
  {https://arxiv.org/abs/https://doi.org/10.1021/acsnano.9b06992}
  {https://doi.org/10.1021/acsnano.9b06992} \BibitemShut {NoStop}%
\bibitem [{\citenamefont {Desai}\ \emph {et~al.}(2016)\citenamefont {Desai},
  \citenamefont {Madhvapathy}, \citenamefont {Amani}, \citenamefont {Kiriya},
  \citenamefont {Hettick}, \citenamefont {Tosun}, \citenamefont {Zhou},
  \citenamefont {Dubey}, \citenamefont {Ager~III}, \citenamefont {Chrzan},\
  and\ \citenamefont {Javey}}]{Au_2Dexfoliation_AdvMatter2016}%
  \BibitemOpen
  \bibfield  {author} {\bibinfo {author} {\bibfnamefont {S.~B.}\ \bibnamefont
  {Desai}}, \bibinfo {author} {\bibfnamefont {S.~R.}\ \bibnamefont
  {Madhvapathy}}, \bibinfo {author} {\bibfnamefont {M.}~\bibnamefont {Amani}},
  \bibinfo {author} {\bibfnamefont {D.}~\bibnamefont {Kiriya}}, \bibinfo
  {author} {\bibfnamefont {M.}~\bibnamefont {Hettick}}, \bibinfo {author}
  {\bibfnamefont {M.}~\bibnamefont {Tosun}}, \bibinfo {author} {\bibfnamefont
  {Y.}~\bibnamefont {Zhou}}, \bibinfo {author} {\bibfnamefont {M.}~\bibnamefont
  {Dubey}}, \bibinfo {author} {\bibfnamefont {J.~W.}\ \bibnamefont {Ager~III}},
  \bibinfo {author} {\bibfnamefont {D.}~\bibnamefont {Chrzan}},\ and\ \bibinfo
  {author} {\bibfnamefont {A.}~\bibnamefont {Javey}},\ }\bibfield  {title}
  {\bibinfo {title} {Gold-mediated exfoliation of ultralarge
  optoelectronically-perfect monolayers},\ }\href
  {https://doi.org/https://doi.org/10.1002/adma.201506171} {\bibfield
  {journal} {\bibinfo  {journal} {Advanced Materials}\ }\textbf {\bibinfo
  {volume} {28}},\ \bibinfo {pages} {4053} (\bibinfo {year} {2016})},\ \Eprint
  {https://arxiv.org/abs/https://onlinelibrary.wiley.com/doi/pdf/10.1002/adma.201506171}
  {https://onlinelibrary.wiley.com/doi/pdf/10.1002/adma.201506171} \BibitemShut
  {NoStop}%
\bibitem [{\citenamefont {Velický}\ \emph {et~al.}(2018)\citenamefont
  {Velický}, \citenamefont {Donnelly}, \citenamefont {Hendren}, \citenamefont
  {McFarland}, \citenamefont {Scullion}, \citenamefont {DeBenedetti},
  \citenamefont {Correa}, \citenamefont {Han}, \citenamefont {Wain},
  \citenamefont {Hines}, \citenamefont {Muller}, \citenamefont {Novoselov},
  \citenamefont {Abruña}, \citenamefont {Bowman}, \citenamefont {Santos},\
  and\ \citenamefont {Huang}}]{Au_2D_MoS2_Exfol_ACSnano2018}%
  \BibitemOpen
  \bibfield  {author} {\bibinfo {author} {\bibfnamefont {M.}~\bibnamefont
  {Velický}}, \bibinfo {author} {\bibfnamefont {G.~E.}\ \bibnamefont
  {Donnelly}}, \bibinfo {author} {\bibfnamefont {W.~R.}\ \bibnamefont
  {Hendren}}, \bibinfo {author} {\bibfnamefont {S.}~\bibnamefont {McFarland}},
  \bibinfo {author} {\bibfnamefont {D.}~\bibnamefont {Scullion}}, \bibinfo
  {author} {\bibfnamefont {W.~J.~I.}\ \bibnamefont {DeBenedetti}}, \bibinfo
  {author} {\bibfnamefont {G.~C.}\ \bibnamefont {Correa}}, \bibinfo {author}
  {\bibfnamefont {Y.}~\bibnamefont {Han}}, \bibinfo {author} {\bibfnamefont
  {A.~J.}\ \bibnamefont {Wain}}, \bibinfo {author} {\bibfnamefont {M.~A.}\
  \bibnamefont {Hines}}, \bibinfo {author} {\bibfnamefont {D.~A.}\ \bibnamefont
  {Muller}}, \bibinfo {author} {\bibfnamefont {K.~S.}\ \bibnamefont
  {Novoselov}}, \bibinfo {author} {\bibfnamefont {H.~D.}\ \bibnamefont
  {Abruña}}, \bibinfo {author} {\bibfnamefont {R.~M.}\ \bibnamefont {Bowman}},
  \bibinfo {author} {\bibfnamefont {E.~J.~G.}\ \bibnamefont {Santos}},\ and\
  \bibinfo {author} {\bibfnamefont {F.}~\bibnamefont {Huang}},\ }\bibfield
  {title} {\bibinfo {title} {Mechanism of gold-assisted exfoliation of
  centimeter-sized transition-metal dichalcogenide monolayers},\ }\href
  {https://doi.org/10.1021/acsnano.8b06101} {\bibfield  {journal} {\bibinfo
  {journal} {ACS Nano}\ }\textbf {\bibinfo {volume} {12}},\ \bibinfo {pages}
  {10463} (\bibinfo {year} {2018})},\ \Eprint
  {https://arxiv.org/abs/https://doi.org/10.1021/acsnano.8b06101}
  {https://doi.org/10.1021/acsnano.8b06101} \BibitemShut {NoStop}%
\bibitem [{\citenamefont {Liu}\ \emph {et~al.}(2020)\citenamefont {Liu},
  \citenamefont {Wu}, \citenamefont {Bai}, \citenamefont {Chae}, \citenamefont
  {Li}, \citenamefont {Wang}, \citenamefont {Hone},\ and\ \citenamefont
  {Zhu}}]{Au_2Dexfoliation_Science2020}%
  \BibitemOpen
  \bibfield  {author} {\bibinfo {author} {\bibfnamefont {F.}~\bibnamefont
  {Liu}}, \bibinfo {author} {\bibfnamefont {W.}~\bibnamefont {Wu}}, \bibinfo
  {author} {\bibfnamefont {Y.}~\bibnamefont {Bai}}, \bibinfo {author}
  {\bibfnamefont {S.~H.}\ \bibnamefont {Chae}}, \bibinfo {author}
  {\bibfnamefont {Q.}~\bibnamefont {Li}}, \bibinfo {author} {\bibfnamefont
  {J.}~\bibnamefont {Wang}}, \bibinfo {author} {\bibfnamefont {J.}~\bibnamefont
  {Hone}},\ and\ \bibinfo {author} {\bibfnamefont {X.-Y.}\ \bibnamefont
  {Zhu}},\ }\bibfield  {title} {\bibinfo {title} {Disassembling 2{D} van der
  {W}aals crystals into macroscopic monolayers and reassembling into artificial
  lattices},\ }\href {https://doi.org/10.1126/science.aba1416} {\bibfield
  {journal} {\bibinfo  {journal} {Science}\ }\textbf {\bibinfo {volume}
  {367}},\ \bibinfo {pages} {903} (\bibinfo {year} {2020})},\ \Eprint
  {https://arxiv.org/abs/https://www.science.org/doi/pdf/10.1126/science.aba1416}
  {https://www.science.org/doi/pdf/10.1126/science.aba1416} \BibitemShut
  {NoStop}%
\bibitem [{\citenamefont {Huang}\ \emph {et~al.}(2020)\citenamefont {Huang},
  \citenamefont {Pan}, \citenamefont {Yang}, \citenamefont {Bao}, \citenamefont
  {Meng}, \citenamefont {Luo}, \citenamefont {Cai}, \citenamefont {Liu},
  \citenamefont {Zhao}, \citenamefont {Zhou}, \citenamefont {Wu}, \citenamefont
  {Zhu}, \citenamefont {Huang}, \citenamefont {Liu}, \citenamefont {Liu},
  \citenamefont {Cheng}, \citenamefont {Wu}, \citenamefont {Tian},
  \citenamefont {Gu}, \citenamefont {Shi}, \citenamefont {Guo}, \citenamefont
  {Cheng}, \citenamefont {Hu}, \citenamefont {Zhao}, \citenamefont {Yang},
  \citenamefont {Sutter}, \citenamefont {Sutter}, \citenamefont {Wang},
  \citenamefont {Ji}, \citenamefont {Zhou},\ and\ \citenamefont
  {Gao}}]{Au_assisted_2Dexfoliation_Nture2020}%
  \BibitemOpen
  \bibfield  {author} {\bibinfo {author} {\bibfnamefont {Y.}~\bibnamefont
  {Huang}}, \bibinfo {author} {\bibfnamefont {Y.-H.}\ \bibnamefont {Pan}},
  \bibinfo {author} {\bibfnamefont {R.}~\bibnamefont {Yang}}, \bibinfo {author}
  {\bibfnamefont {L.-H.}\ \bibnamefont {Bao}}, \bibinfo {author} {\bibfnamefont
  {L.}~\bibnamefont {Meng}}, \bibinfo {author} {\bibfnamefont {H.-L.}\
  \bibnamefont {Luo}}, \bibinfo {author} {\bibfnamefont {Y.-Q.}\ \bibnamefont
  {Cai}}, \bibinfo {author} {\bibfnamefont {G.-D.}\ \bibnamefont {Liu}},
  \bibinfo {author} {\bibfnamefont {W.-J.}\ \bibnamefont {Zhao}}, \bibinfo
  {author} {\bibfnamefont {Z.}~\bibnamefont {Zhou}}, \bibinfo {author}
  {\bibfnamefont {L.-M.}\ \bibnamefont {Wu}}, \bibinfo {author} {\bibfnamefont
  {Z.-L.}\ \bibnamefont {Zhu}}, \bibinfo {author} {\bibfnamefont
  {M.}~\bibnamefont {Huang}}, \bibinfo {author} {\bibfnamefont {L.-W.}\
  \bibnamefont {Liu}}, \bibinfo {author} {\bibfnamefont {L.}~\bibnamefont
  {Liu}}, \bibinfo {author} {\bibfnamefont {P.}~\bibnamefont {Cheng}}, \bibinfo
  {author} {\bibfnamefont {K.-H.}\ \bibnamefont {Wu}}, \bibinfo {author}
  {\bibfnamefont {S.-B.}\ \bibnamefont {Tian}}, \bibinfo {author}
  {\bibfnamefont {C.-Z.}\ \bibnamefont {Gu}}, \bibinfo {author} {\bibfnamefont
  {Y.-G.}\ \bibnamefont {Shi}}, \bibinfo {author} {\bibfnamefont {Y.-F.}\
  \bibnamefont {Guo}}, \bibinfo {author} {\bibfnamefont {Z.~G.}\ \bibnamefont
  {Cheng}}, \bibinfo {author} {\bibfnamefont {J.-P.}\ \bibnamefont {Hu}},
  \bibinfo {author} {\bibfnamefont {L.}~\bibnamefont {Zhao}}, \bibinfo {author}
  {\bibfnamefont {G.-H.}\ \bibnamefont {Yang}}, \bibinfo {author}
  {\bibfnamefont {E.}~\bibnamefont {Sutter}}, \bibinfo {author} {\bibfnamefont
  {P.}~\bibnamefont {Sutter}}, \bibinfo {author} {\bibfnamefont {Y.-L.}\
  \bibnamefont {Wang}}, \bibinfo {author} {\bibfnamefont {W.}~\bibnamefont
  {Ji}}, \bibinfo {author} {\bibfnamefont {X.-J.}\ \bibnamefont {Zhou}},\ and\
  \bibinfo {author} {\bibfnamefont {H.-J.}\ \bibnamefont {Gao}},\ }\bibfield
  {title} {\bibinfo {title} {Universal mechanical exfoliation of large-area
  2{D} crystals},\ }\href {https://doi.org/10.1038/s41467-020-16266-w}
  {\bibfield  {journal} {\bibinfo  {journal} {Nature Communications}\ }\textbf
  {\bibinfo {volume} {11}},\ \bibinfo {pages} {2453} (\bibinfo {year}
  {2020})},\ \Eprint
  {https://arxiv.org/abs/https://doi.org/10.1038/s41467-020-16266-w}
  {https://doi.org/10.1038/s41467-020-16266-w} \BibitemShut {NoStop}%
\bibitem [{\citenamefont {Heyl}\ \emph {et~al.}(2020)\citenamefont {Heyl},
  \citenamefont {Burmeister}, \citenamefont {Schultz}, \citenamefont
  {Pallasch}, \citenamefont {Ligorio}, \citenamefont {Koch},\ and\
  \citenamefont {List-Kratochvil}}]{Au_ThermallyActivatedTransfer_Koch2020}%
  \BibitemOpen
  \bibfield  {author} {\bibinfo {author} {\bibfnamefont {M.}~\bibnamefont
  {Heyl}}, \bibinfo {author} {\bibfnamefont {D.}~\bibnamefont {Burmeister}},
  \bibinfo {author} {\bibfnamefont {T.}~\bibnamefont {Schultz}}, \bibinfo
  {author} {\bibfnamefont {S.}~\bibnamefont {Pallasch}}, \bibinfo {author}
  {\bibfnamefont {G.}~\bibnamefont {Ligorio}}, \bibinfo {author} {\bibfnamefont
  {N.}~\bibnamefont {Koch}},\ and\ \bibinfo {author} {\bibfnamefont {E.~J.~W.}\
  \bibnamefont {List-Kratochvil}},\ }\bibfield  {title} {\bibinfo {title}
  {Thermally activated gold-mediated transition metal dichalcogenide
  exfoliation and a unique gold-mediated transfer},\ }\href
  {https://doi.org/https://doi.org/10.1002/pssr.202000408} {\bibfield
  {journal} {\bibinfo  {journal} {physica status solidi (RRL) – Rapid
  Research Letters}\ }\textbf {\bibinfo {volume} {14}},\ \bibinfo {pages}
  {2000408} (\bibinfo {year} {2020})},\ \Eprint
  {https://arxiv.org/abs/https://onlinelibrary.wiley.com/doi/pdf/10.1002/pssr.202000408}
  {https://onlinelibrary.wiley.com/doi/pdf/10.1002/pssr.202000408} \BibitemShut
  {NoStop}%
\bibitem [{\citenamefont {Li}\ \emph {et~al.}(2015)\citenamefont {Li},
  \citenamefont {Shi}, \citenamefont {Cheng}, \citenamefont {Lu}, \citenamefont
  {Lin}, \citenamefont {Tang}, \citenamefont {Tsai}, \citenamefont {Chu},
  \citenamefont {Wei}, \citenamefont {He}, \citenamefont {Chang}, \citenamefont
  {Suenaga},\ and\ \citenamefont {Li}}]{MoS2_WSe2_Epitaxy_Science2015}%
  \BibitemOpen
  \bibfield  {author} {\bibinfo {author} {\bibfnamefont {M.-Y.}\ \bibnamefont
  {Li}}, \bibinfo {author} {\bibfnamefont {Y.}~\bibnamefont {Shi}}, \bibinfo
  {author} {\bibfnamefont {C.-C.}\ \bibnamefont {Cheng}}, \bibinfo {author}
  {\bibfnamefont {L.-S.}\ \bibnamefont {Lu}}, \bibinfo {author} {\bibfnamefont
  {Y.-C.}\ \bibnamefont {Lin}}, \bibinfo {author} {\bibfnamefont {H.-L.}\
  \bibnamefont {Tang}}, \bibinfo {author} {\bibfnamefont {M.-L.}\ \bibnamefont
  {Tsai}}, \bibinfo {author} {\bibfnamefont {C.-W.}\ \bibnamefont {Chu}},
  \bibinfo {author} {\bibfnamefont {K.-H.}\ \bibnamefont {Wei}}, \bibinfo
  {author} {\bibfnamefont {J.-H.}\ \bibnamefont {He}}, \bibinfo {author}
  {\bibfnamefont {W.-H.}\ \bibnamefont {Chang}}, \bibinfo {author}
  {\bibfnamefont {K.}~\bibnamefont {Suenaga}},\ and\ \bibinfo {author}
  {\bibfnamefont {L.-J.}\ \bibnamefont {Li}},\ }\bibfield  {title} {\bibinfo
  {title} {Epitaxial growth of a monolayer {WS}e$_2$-{M}o{S}$_2$ lateral p-n
  junction with an atomically sharp interface},\ }\href
  {https://doi.org/10.1126/science.aab4097} {\bibfield  {journal} {\bibinfo
  {journal} {Science}\ }\textbf {\bibinfo {volume} {349}},\ \bibinfo {pages}
  {524} (\bibinfo {year} {2015})},\ \Eprint
  {https://arxiv.org/abs/https://www.science.org/doi/pdf/10.1126/science.aab4097}
  {https://www.science.org/doi/pdf/10.1126/science.aab4097} \BibitemShut
  {NoStop}%
\bibitem [{\citenamefont {Han}\ \emph {et~al.}(2019)\citenamefont {Han},
  \citenamefont {Aljarb}, \citenamefont {Liu}, \citenamefont {Li},
  \citenamefont {Ma}, \citenamefont {Xue}, \citenamefont {Lopatin},
  \citenamefont {Yang}, \citenamefont {Huang}, \citenamefont {Wan},
  \citenamefont {Zhang}, \citenamefont {Xiong}, \citenamefont {Huang},
  \citenamefont {Tung}, \citenamefont {Anthopoulos},\ and\ \citenamefont
  {Li}}]{WSe2_BL_sapphire_AFMheight_Li}%
  \BibitemOpen
  \bibfield  {author} {\bibinfo {author} {\bibfnamefont {A.}~\bibnamefont
  {Han}}, \bibinfo {author} {\bibfnamefont {A.}~\bibnamefont {Aljarb}},
  \bibinfo {author} {\bibfnamefont {S.}~\bibnamefont {Liu}}, \bibinfo {author}
  {\bibfnamefont {P.}~\bibnamefont {Li}}, \bibinfo {author} {\bibfnamefont
  {C.}~\bibnamefont {Ma}}, \bibinfo {author} {\bibfnamefont {F.}~\bibnamefont
  {Xue}}, \bibinfo {author} {\bibfnamefont {S.}~\bibnamefont {Lopatin}},
  \bibinfo {author} {\bibfnamefont {C.-W.}\ \bibnamefont {Yang}}, \bibinfo
  {author} {\bibfnamefont {J.-K.}\ \bibnamefont {Huang}}, \bibinfo {author}
  {\bibfnamefont {Y.}~\bibnamefont {Wan}}, \bibinfo {author} {\bibfnamefont
  {X.}~\bibnamefont {Zhang}}, \bibinfo {author} {\bibfnamefont
  {Q.}~\bibnamefont {Xiong}}, \bibinfo {author} {\bibfnamefont {K.-W.}\
  \bibnamefont {Huang}}, \bibinfo {author} {\bibfnamefont {V.}~\bibnamefont
  {Tung}}, \bibinfo {author} {\bibfnamefont {T.~D.}\ \bibnamefont
  {Anthopoulos}},\ and\ \bibinfo {author} {\bibfnamefont {L.-J.}\ \bibnamefont
  {Li}},\ }\bibfield  {title} {\bibinfo {title} {Growth of 2{H} stacked
  {WS}e$_2$ bilayers on sapphire},\ }\href {https://doi.org/10.1039/C9NH00260J}
  {\bibfield  {journal} {\bibinfo  {journal} {Nanoscale Horizons}\ }\textbf
  {\bibinfo {volume} {4}},\ \bibinfo {pages} {1434} (\bibinfo {year}
  {2019})}\BibitemShut {NoStop}%
\bibitem [{\citenamefont {Wilson}\ \emph {et~al.}(2017)\citenamefont {Wilson},
  \citenamefont {Nguyen}, \citenamefont {Seyler}, \citenamefont {Rivera},
  \citenamefont {Marsden}, \citenamefont {Laker}, \citenamefont
  {Constantinescu}, \citenamefont {Kandyba}, \citenamefont {Barinov},
  \citenamefont {Hine}, \citenamefont {Xu},\ and\ \citenamefont
  {Cobden}}]{WSe2_uARPES_ML_BL_bulk}%
  \BibitemOpen
  \bibfield  {author} {\bibinfo {author} {\bibfnamefont {N.~R.}\ \bibnamefont
  {Wilson}}, \bibinfo {author} {\bibfnamefont {P.~V.}\ \bibnamefont {Nguyen}},
  \bibinfo {author} {\bibfnamefont {K.}~\bibnamefont {Seyler}}, \bibinfo
  {author} {\bibfnamefont {P.}~\bibnamefont {Rivera}}, \bibinfo {author}
  {\bibfnamefont {A.~J.}\ \bibnamefont {Marsden}}, \bibinfo {author}
  {\bibfnamefont {Z.~P.~L.}\ \bibnamefont {Laker}}, \bibinfo {author}
  {\bibfnamefont {G.~C.}\ \bibnamefont {Constantinescu}}, \bibinfo {author}
  {\bibfnamefont {V.}~\bibnamefont {Kandyba}}, \bibinfo {author} {\bibfnamefont
  {A.}~\bibnamefont {Barinov}}, \bibinfo {author} {\bibfnamefont {N.~D.~M.}\
  \bibnamefont {Hine}}, \bibinfo {author} {\bibfnamefont {X.}~\bibnamefont
  {Xu}},\ and\ \bibinfo {author} {\bibfnamefont {D.~H.}\ \bibnamefont
  {Cobden}},\ }\bibfield  {title} {\bibinfo {title} {Determination of band
  offsets, hybridization, and exciton binding in 2{D} semiconductor
  heterostructures},\ }\href {https://doi.org/10.1126/sciadv.1601832}
  {\bibfield  {journal} {\bibinfo  {journal} {Science Advances}\ }\textbf
  {\bibinfo {volume} {3}},\ \bibinfo {pages} {e1601832} (\bibinfo {year}
  {2017})},\ \Eprint
  {https://arxiv.org/abs/https://www.science.org/doi/pdf/10.1126/sciadv.1601832}
  {https://www.science.org/doi/pdf/10.1126/sciadv.1601832} \BibitemShut
  {NoStop}%
\bibitem [{\citenamefont {Zhang}\ \emph {et~al.}(2016)\citenamefont {Zhang},
  \citenamefont {Ugeda}, \citenamefont {Jin}, \citenamefont {Shi},
  \citenamefont {Bradley}, \citenamefont {Martín-Recio}, \citenamefont {Ryu},
  \citenamefont {Kim}, \citenamefont {Tang}, \citenamefont {Kim}, \citenamefont
  {Zhou}, \citenamefont {Hwang}, \citenamefont {Chen}, \citenamefont {Wang},
  \citenamefont {Crommie}, \citenamefont {Hussain}, \citenamefont {Shen},\ and\
  \citenamefont {Mo}}]{WSe2_ThinFilmsMBE_SKMo}%
  \BibitemOpen
  \bibfield  {author} {\bibinfo {author} {\bibfnamefont {Y.}~\bibnamefont
  {Zhang}}, \bibinfo {author} {\bibfnamefont {M.~M.}\ \bibnamefont {Ugeda}},
  \bibinfo {author} {\bibfnamefont {C.}~\bibnamefont {Jin}}, \bibinfo {author}
  {\bibfnamefont {S.-F.}\ \bibnamefont {Shi}}, \bibinfo {author} {\bibfnamefont
  {A.~J.}\ \bibnamefont {Bradley}}, \bibinfo {author} {\bibfnamefont
  {A.}~\bibnamefont {Martín-Recio}}, \bibinfo {author} {\bibfnamefont
  {H.}~\bibnamefont {Ryu}}, \bibinfo {author} {\bibfnamefont {J.}~\bibnamefont
  {Kim}}, \bibinfo {author} {\bibfnamefont {S.}~\bibnamefont {Tang}}, \bibinfo
  {author} {\bibfnamefont {Y.}~\bibnamefont {Kim}}, \bibinfo {author}
  {\bibfnamefont {B.}~\bibnamefont {Zhou}}, \bibinfo {author} {\bibfnamefont
  {C.}~\bibnamefont {Hwang}}, \bibinfo {author} {\bibfnamefont
  {Y.}~\bibnamefont {Chen}}, \bibinfo {author} {\bibfnamefont {F.}~\bibnamefont
  {Wang}}, \bibinfo {author} {\bibfnamefont {M.~F.}\ \bibnamefont {Crommie}},
  \bibinfo {author} {\bibfnamefont {Z.}~\bibnamefont {Hussain}}, \bibinfo
  {author} {\bibfnamefont {Z.-X.}\ \bibnamefont {Shen}},\ and\ \bibinfo
  {author} {\bibfnamefont {S.-K.}\ \bibnamefont {Mo}},\ }\bibfield  {title}
  {\bibinfo {title} {Electronic structure, surface doping, and optical response
  in epitaxial {WSe}$_2$ thin films},\ }\href
  {https://doi.org/10.1021/acs.nanolett.6b00059} {\bibfield  {journal}
  {\bibinfo  {journal} {Nano Letters}\ }\textbf {\bibinfo {volume} {16}},\
  \bibinfo {pages} {2485} (\bibinfo {year} {2016})},\ \Eprint
  {https://arxiv.org/abs/https://doi.org/10.1021/acs.nanolett.6b00059}
  {https://doi.org/10.1021/acs.nanolett.6b00059} \BibitemShut {NoStop}%
\bibitem [{\citenamefont {Bruix}\ \emph {et~al.}(2016)\citenamefont {Bruix},
  \citenamefont {Miwa}, \citenamefont {Hauptmann}, \citenamefont {Wegner},
  \citenamefont {Ulstrup}, \citenamefont {Grønborg}, \citenamefont {Sanders},
  \citenamefont {Dendzik}, \citenamefont {Grubišić~Čabo}, \citenamefont
  {Bianchi}, \citenamefont {Lauritsen}, \citenamefont {Khajetoorians},
  \citenamefont {Hammer},\ and\ \citenamefont {Hofmann}}]{Bruix16}%
  \BibitemOpen
  \bibfield  {author} {\bibinfo {author} {\bibfnamefont {A.}~\bibnamefont
  {Bruix}}, \bibinfo {author} {\bibfnamefont {J.~A.}\ \bibnamefont {Miwa}},
  \bibinfo {author} {\bibfnamefont {N.}~\bibnamefont {Hauptmann}}, \bibinfo
  {author} {\bibfnamefont {D.}~\bibnamefont {Wegner}}, \bibinfo {author}
  {\bibfnamefont {S.}~\bibnamefont {Ulstrup}}, \bibinfo {author} {\bibfnamefont
  {S.~S.}\ \bibnamefont {Grønborg}}, \bibinfo {author} {\bibfnamefont {C.~E.}\
  \bibnamefont {Sanders}}, \bibinfo {author} {\bibfnamefont {M.}~\bibnamefont
  {Dendzik}}, \bibinfo {author} {\bibfnamefont {A.}~\bibnamefont
  {Grubišić~Čabo}}, \bibinfo {author} {\bibfnamefont {M.}~\bibnamefont
  {Bianchi}}, \bibinfo {author} {\bibfnamefont {J.~V.}\ \bibnamefont
  {Lauritsen}}, \bibinfo {author} {\bibfnamefont {A.~A.}\ \bibnamefont
  {Khajetoorians}}, \bibinfo {author} {\bibfnamefont {B.}~\bibnamefont
  {Hammer}},\ and\ \bibinfo {author} {\bibfnamefont {P.}~\bibnamefont
  {Hofmann}},\ }\bibfield  {title} {\bibinfo {title} {Single-layer {M}o{S}$_2$
  on {A}u(111): Band gap renormalization and substrate interaction},\ }\href
  {https://doi.org/10.1103/PhysRevB.93.165422} {\bibfield  {journal} {\bibinfo
  {journal} {Physical Review B}\ }\textbf {\bibinfo {volume} {93}},\ \bibinfo
  {pages} {165422} (\bibinfo {year} {2016})}\BibitemShut {NoStop}%
\bibitem [{\citenamefont {Dendzik}\ \emph {et~al.}(2017)\citenamefont
  {Dendzik}, \citenamefont {Bruix}, \citenamefont {Michiardi}, \citenamefont
  {Ngankeu}, \citenamefont {Bianchi}, \citenamefont {Miwa}, \citenamefont
  {Hammer}, \citenamefont {Hofmann},\ and\ \citenamefont
  {Sanders}}]{WS2_Ag_metalTrans_Mac}%
  \BibitemOpen
  \bibfield  {author} {\bibinfo {author} {\bibfnamefont {M.}~\bibnamefont
  {Dendzik}}, \bibinfo {author} {\bibfnamefont {A.}~\bibnamefont {Bruix}},
  \bibinfo {author} {\bibfnamefont {M.}~\bibnamefont {Michiardi}}, \bibinfo
  {author} {\bibfnamefont {A.~S.}\ \bibnamefont {Ngankeu}}, \bibinfo {author}
  {\bibfnamefont {M.}~\bibnamefont {Bianchi}}, \bibinfo {author} {\bibfnamefont
  {J.~A.}\ \bibnamefont {Miwa}}, \bibinfo {author} {\bibfnamefont
  {B.}~\bibnamefont {Hammer}}, \bibinfo {author} {\bibfnamefont
  {P.}~\bibnamefont {Hofmann}},\ and\ \bibinfo {author} {\bibfnamefont {C.~E.}\
  \bibnamefont {Sanders}},\ }\bibfield  {title} {\bibinfo {title}
  {Substrate-induced semiconductor-to-metal transition in monolayer {WS}$_2$},\
  }\href {https://doi.org/10.1103/PhysRevB.96.235440} {\bibfield  {journal}
  {\bibinfo  {journal} {Phys. Rev. B}\ }\textbf {\bibinfo {volume} {96}},\
  \bibinfo {pages} {235440} (\bibinfo {year} {2017})}\BibitemShut {NoStop}%
\bibitem [{\citenamefont {Dendzik}\ \emph {et~al.}(2015)\citenamefont
  {Dendzik}, \citenamefont {Michiardi}, \citenamefont {Sanders}, \citenamefont
  {Bianchi}, \citenamefont {Miwa}, \citenamefont {Gr\o{}nborg}, \citenamefont
  {Lauritsen}, \citenamefont {Bruix}, \citenamefont {Hammer},\ and\
  \citenamefont {Hofmann}}]{WS2GrowthAu111_VBsplit}%
  \BibitemOpen
  \bibfield  {author} {\bibinfo {author} {\bibfnamefont {M.}~\bibnamefont
  {Dendzik}}, \bibinfo {author} {\bibfnamefont {M.}~\bibnamefont {Michiardi}},
  \bibinfo {author} {\bibfnamefont {C.}~\bibnamefont {Sanders}}, \bibinfo
  {author} {\bibfnamefont {M.}~\bibnamefont {Bianchi}}, \bibinfo {author}
  {\bibfnamefont {J.~A.}\ \bibnamefont {Miwa}}, \bibinfo {author}
  {\bibfnamefont {S.~S.}\ \bibnamefont {Gr\o{}nborg}}, \bibinfo {author}
  {\bibfnamefont {J.~V.}\ \bibnamefont {Lauritsen}}, \bibinfo {author}
  {\bibfnamefont {A.}~\bibnamefont {Bruix}}, \bibinfo {author} {\bibfnamefont
  {B.}~\bibnamefont {Hammer}},\ and\ \bibinfo {author} {\bibfnamefont
  {P.}~\bibnamefont {Hofmann}},\ }\bibfield  {title} {\bibinfo {title} {Growth
  and electronic structure of epitaxial single-layer {WS}$_2$ on {A}u(111)},\
  }\href {https://doi.org/10.1103/PhysRevB.92.245442} {\bibfield  {journal}
  {\bibinfo  {journal} {Phys. Rev. B}\ }\textbf {\bibinfo {volume} {92}},\
  \bibinfo {pages} {245442} (\bibinfo {year} {2015})}\BibitemShut {NoStop}%
\bibitem [{\citenamefont {Ulstrup}\ \emph {et~al.}(2017)\citenamefont
  {Ulstrup}, \citenamefont {\ifmmode~\check{C}\else \v{C}\fi{}abo},
  \citenamefont {Biswas}, \citenamefont {Riley}, \citenamefont {Dendzik},
  \citenamefont {Sanders}, \citenamefont {Bianchi}, \citenamefont {Cacho},
  \citenamefont {Matselyukh}, \citenamefont {Chapman}, \citenamefont
  {Springate}, \citenamefont {King}, \citenamefont {Miwa},\ and\ \citenamefont
  {Hofmann}}]{WS2Ag111_TRARPES_VBsplit}%
  \BibitemOpen
  \bibfield  {author} {\bibinfo {author} {\bibfnamefont {S.}~\bibnamefont
  {Ulstrup}}, \bibinfo {author} {\bibfnamefont {A.~G. c. v. a.~c.}\
  \bibnamefont {\ifmmode~\check{C}\else \v{C}\fi{}abo}}, \bibinfo {author}
  {\bibfnamefont {D.}~\bibnamefont {Biswas}}, \bibinfo {author} {\bibfnamefont
  {J.~M.}\ \bibnamefont {Riley}}, \bibinfo {author} {\bibfnamefont
  {M.}~\bibnamefont {Dendzik}}, \bibinfo {author} {\bibfnamefont {C.~E.}\
  \bibnamefont {Sanders}}, \bibinfo {author} {\bibfnamefont {M.}~\bibnamefont
  {Bianchi}}, \bibinfo {author} {\bibfnamefont {C.}~\bibnamefont {Cacho}},
  \bibinfo {author} {\bibfnamefont {D.}~\bibnamefont {Matselyukh}}, \bibinfo
  {author} {\bibfnamefont {R.~T.}\ \bibnamefont {Chapman}}, \bibinfo {author}
  {\bibfnamefont {E.}~\bibnamefont {Springate}}, \bibinfo {author}
  {\bibfnamefont {P.~D.~C.}\ \bibnamefont {King}}, \bibinfo {author}
  {\bibfnamefont {J.~A.}\ \bibnamefont {Miwa}},\ and\ \bibinfo {author}
  {\bibfnamefont {P.}~\bibnamefont {Hofmann}},\ }\bibfield  {title} {\bibinfo
  {title} {Spin and valley control of free carriers in single-layer {WS}$_2$},\
  }\href {https://doi.org/10.1103/PhysRevB.95.041405} {\bibfield  {journal}
  {\bibinfo  {journal} {Phys. Rev. B}\ }\textbf {\bibinfo {volume} {95}},\
  \bibinfo {pages} {041405} (\bibinfo {year} {2017})}\BibitemShut {NoStop}%
\bibitem [{\citenamefont {Reinert}\ \emph {et~al.}(2001)\citenamefont
  {Reinert}, \citenamefont {Nicolay}, \citenamefont {Schmidt}, \citenamefont
  {Ehm},\ and\ \citenamefont {H\"ufner}}]{111NobleMetalSS}%
  \BibitemOpen
  \bibfield  {author} {\bibinfo {author} {\bibfnamefont {F.}~\bibnamefont
  {Reinert}}, \bibinfo {author} {\bibfnamefont {G.}~\bibnamefont {Nicolay}},
  \bibinfo {author} {\bibfnamefont {S.}~\bibnamefont {Schmidt}}, \bibinfo
  {author} {\bibfnamefont {D.}~\bibnamefont {Ehm}},\ and\ \bibinfo {author}
  {\bibfnamefont {S.}~\bibnamefont {H\"ufner}},\ }\bibfield  {title} {\bibinfo
  {title} {Direct measurements of the l-gap surface states on the (111) face of
  noble metals by photoelectron spectroscopy},\ }\href
  {https://doi.org/10.1103/PhysRevB.63.115415} {\bibfield  {journal} {\bibinfo
  {journal} {Phys. Rev. B}\ }\textbf {\bibinfo {volume} {63}},\ \bibinfo
  {pages} {115415} (\bibinfo {year} {2001})}\BibitemShut {NoStop}%
\bibitem [{\citenamefont {Yokoya}\ \emph {et~al.}(2001)\citenamefont {Yokoya},
  \citenamefont {Kiss}, \citenamefont {Chainani}, \citenamefont {Shin},
  \citenamefont {Nohara},\ and\ \citenamefont {Takagi}}]{NbSe2_bulk_Supercond}%
  \BibitemOpen
  \bibfield  {author} {\bibinfo {author} {\bibfnamefont {T.}~\bibnamefont
  {Yokoya}}, \bibinfo {author} {\bibfnamefont {T.}~\bibnamefont {Kiss}},
  \bibinfo {author} {\bibfnamefont {A.}~\bibnamefont {Chainani}}, \bibinfo
  {author} {\bibfnamefont {S.}~\bibnamefont {Shin}}, \bibinfo {author}
  {\bibfnamefont {M.}~\bibnamefont {Nohara}},\ and\ \bibinfo {author}
  {\bibfnamefont {H.}~\bibnamefont {Takagi}},\ }\bibfield  {title} {\bibinfo
  {title} {Fermi surface sheet-dependent superconductivity in {2H-NbSe$_2$}},\
  }\href {https://doi.org/10.1126/science.1065068} {\bibfield  {journal}
  {\bibinfo  {journal} {Science}\ }\textbf {\bibinfo {volume} {294}},\ \bibinfo
  {pages} {2518} (\bibinfo {year} {2001})},\ \Eprint
  {https://arxiv.org/abs/https://www.science.org/doi/pdf/10.1126/science.1065068}
  {https://www.science.org/doi/pdf/10.1126/science.1065068} \BibitemShut
  {NoStop}%
\bibitem [{\citenamefont {Wu}\ and\ \citenamefont
  {Lieber}(1989)}]{TaS2_CDW_1989}%
  \BibitemOpen
  \bibfield  {author} {\bibinfo {author} {\bibfnamefont {X.~L.}\ \bibnamefont
  {Wu}}\ and\ \bibinfo {author} {\bibfnamefont {C.~M.}\ \bibnamefont
  {Lieber}},\ }\bibfield  {title} {\bibinfo {title} {Hexagonal domain-like
  charge density wave phase of {TaS$_2$} determined by scanning tunneling
  microscopy},\ }\href {https://doi.org/10.1126/science.243.4899.1703}
  {\bibfield  {journal} {\bibinfo  {journal} {Science}\ }\textbf {\bibinfo
  {volume} {243}},\ \bibinfo {pages} {1703} (\bibinfo {year} {1989})},\ \Eprint
  {https://arxiv.org/abs/https://www.science.org/doi/pdf/10.1126/science.243.4899.1703}
  {https://www.science.org/doi/pdf/10.1126/science.243.4899.1703} \BibitemShut
  {NoStop}%
\bibitem [{\citenamefont {Stahl}\ \emph {et~al.}(2020)\citenamefont {Stahl},
  \citenamefont {Kusch}, \citenamefont {Heinsch}, \citenamefont {Garbarino},
  \citenamefont {Kretzschmar}, \citenamefont {Hanff}, \citenamefont
  {Rossnagel}, \citenamefont {Geck},\ and\ \citenamefont
  {Ritschel}}]{TaS2_CDW_Rossnagel}%
  \BibitemOpen
  \bibfield  {author} {\bibinfo {author} {\bibfnamefont {Q.}~\bibnamefont
  {Stahl}}, \bibinfo {author} {\bibfnamefont {M.}~\bibnamefont {Kusch}},
  \bibinfo {author} {\bibfnamefont {F.}~\bibnamefont {Heinsch}}, \bibinfo
  {author} {\bibfnamefont {G.}~\bibnamefont {Garbarino}}, \bibinfo {author}
  {\bibfnamefont {N.}~\bibnamefont {Kretzschmar}}, \bibinfo {author}
  {\bibfnamefont {K.}~\bibnamefont {Hanff}}, \bibinfo {author} {\bibfnamefont
  {K.}~\bibnamefont {Rossnagel}}, \bibinfo {author} {\bibfnamefont
  {J.}~\bibnamefont {Geck}},\ and\ \bibinfo {author} {\bibfnamefont
  {T.}~\bibnamefont {Ritschel}},\ }\bibfield  {title} {\bibinfo {title}
  {Collapse of layer dimerization in the photo-induced hidden state of
  {1T-TaS$_2$}},\ }\href {https://doi.org/10.1038/s41467-020-15079-1}
  {\bibfield  {journal} {\bibinfo  {journal} {Nature Communications}\ }\textbf
  {\bibinfo {volume} {11}},\ \bibinfo {pages} {1247} (\bibinfo {year}
  {2020})}\BibitemShut {NoStop}%
\bibitem [{\citenamefont {Shi}\ \emph {et~al.}(2019)\citenamefont {Shi},
  \citenamefont {Kahn}, \citenamefont {Niu}, \citenamefont {Fei}, \citenamefont
  {Sun}, \citenamefont {Cai}, \citenamefont {Francisco}, \citenamefont {Wu},
  \citenamefont {Shen}, \citenamefont {Xu}, \citenamefont {Cobden},\ and\
  \citenamefont {Cui}}]{WTe2_QSHE_realSpace}%
  \BibitemOpen
  \bibfield  {author} {\bibinfo {author} {\bibfnamefont {Y.}~\bibnamefont
  {Shi}}, \bibinfo {author} {\bibfnamefont {J.}~\bibnamefont {Kahn}}, \bibinfo
  {author} {\bibfnamefont {B.}~\bibnamefont {Niu}}, \bibinfo {author}
  {\bibfnamefont {Z.}~\bibnamefont {Fei}}, \bibinfo {author} {\bibfnamefont
  {B.}~\bibnamefont {Sun}}, \bibinfo {author} {\bibfnamefont {X.}~\bibnamefont
  {Cai}}, \bibinfo {author} {\bibfnamefont {B.~A.}\ \bibnamefont {Francisco}},
  \bibinfo {author} {\bibfnamefont {D.}~\bibnamefont {Wu}}, \bibinfo {author}
  {\bibfnamefont {Z.-X.}\ \bibnamefont {Shen}}, \bibinfo {author}
  {\bibfnamefont {X.}~\bibnamefont {Xu}}, \bibinfo {author} {\bibfnamefont
  {D.~H.}\ \bibnamefont {Cobden}},\ and\ \bibinfo {author} {\bibfnamefont
  {Y.-T.}\ \bibnamefont {Cui}},\ }\bibfield  {title} {\bibinfo {title} {Imaging
  quantum spin hall edges in monolayer wte<sub>2</sub>},\ }\href
  {https://doi.org/10.1126/sciadv.aat8799} {\bibfield  {journal} {\bibinfo
  {journal} {Science Advances}\ }\textbf {\bibinfo {volume} {5}},\ \bibinfo
  {pages} {eaat8799} (\bibinfo {year} {2019})},\ \Eprint
  {https://arxiv.org/abs/https://www.science.org/doi/pdf/10.1126/sciadv.aat8799}
  {https://www.science.org/doi/pdf/10.1126/sciadv.aat8799} \BibitemShut
  {NoStop}%
\bibitem [{\citenamefont {Ye}\ \emph {et~al.}(2016)\citenamefont {Ye},
  \citenamefont {Lee}, \citenamefont {Hu}, \citenamefont {Mao}, \citenamefont
  {Wei},\ and\ \citenamefont {Feng}}]{WTe2_exfoliation_ambient_unstable}%
  \BibitemOpen
  \bibfield  {author} {\bibinfo {author} {\bibfnamefont {F.}~\bibnamefont
  {Ye}}, \bibinfo {author} {\bibfnamefont {J.}~\bibnamefont {Lee}}, \bibinfo
  {author} {\bibfnamefont {J.}~\bibnamefont {Hu}}, \bibinfo {author}
  {\bibfnamefont {Z.}~\bibnamefont {Mao}}, \bibinfo {author} {\bibfnamefont
  {J.}~\bibnamefont {Wei}},\ and\ \bibinfo {author} {\bibfnamefont {P.~X.-L.}\
  \bibnamefont {Feng}},\ }\bibfield  {title} {\bibinfo {title} {Environmental
  instability and degradation of single- and few-layer {WTe$_2$} nanosheets in
  ambient conditions},\ }\href
  {https://doi.org/https://doi.org/10.1002/smll.201601207} {\bibfield
  {journal} {\bibinfo  {journal} {Small}\ }\textbf {\bibinfo {volume} {12}},\
  \bibinfo {pages} {5802} (\bibinfo {year} {2016})},\ \Eprint
  {https://arxiv.org/abs/https://onlinelibrary.wiley.com/doi/pdf/10.1002/smll.201601207}
  {https://onlinelibrary.wiley.com/doi/pdf/10.1002/smll.201601207} \BibitemShut
  {NoStop}%
\bibitem [{\citenamefont {Kim}\ \emph {et~al.}(2016)\citenamefont {Kim},
  \citenamefont {Jhon}, \citenamefont {Park}, \citenamefont {Kim},
  \citenamefont {Lee},\ and\ \citenamefont {Jhon}}]{WTe2_Raman_MLtoBulk}%
  \BibitemOpen
  \bibfield  {author} {\bibinfo {author} {\bibfnamefont {Y.}~\bibnamefont
  {Kim}}, \bibinfo {author} {\bibfnamefont {Y.~I.}\ \bibnamefont {Jhon}},
  \bibinfo {author} {\bibfnamefont {J.}~\bibnamefont {Park}}, \bibinfo {author}
  {\bibfnamefont {J.~H.}\ \bibnamefont {Kim}}, \bibinfo {author} {\bibfnamefont
  {S.}~\bibnamefont {Lee}},\ and\ \bibinfo {author} {\bibfnamefont {Y.~M.}\
  \bibnamefont {Jhon}},\ }\bibfield  {title} {\bibinfo {title} {Anomalous raman
  scattering and lattice dynamics in mono- and few-layer wte2},\ }\href
  {https://doi.org/10.1039/C5NR06098B} {\bibfield  {journal} {\bibinfo
  {journal} {Nanoscale}\ }\textbf {\bibinfo {volume} {8}},\ \bibinfo {pages}
  {2309} (\bibinfo {year} {2016})}\BibitemShut {NoStop}%
\bibitem [{\citenamefont {Vogt}\ \emph {et~al.}(2012)\citenamefont {Vogt},
  \citenamefont {De~Padova}, \citenamefont {Quaresima}, \citenamefont {Avila},
  \citenamefont {Frantzeskakis}, \citenamefont {Asensio}, \citenamefont
  {Resta}, \citenamefont {Ealet},\ and\ \citenamefont
  {Le~Lay}}]{Silicene_LeLay2012}%
  \BibitemOpen
  \bibfield  {author} {\bibinfo {author} {\bibfnamefont {P.}~\bibnamefont
  {Vogt}}, \bibinfo {author} {\bibfnamefont {P.}~\bibnamefont {De~Padova}},
  \bibinfo {author} {\bibfnamefont {C.}~\bibnamefont {Quaresima}}, \bibinfo
  {author} {\bibfnamefont {J.}~\bibnamefont {Avila}}, \bibinfo {author}
  {\bibfnamefont {E.}~\bibnamefont {Frantzeskakis}}, \bibinfo {author}
  {\bibfnamefont {M.~C.}\ \bibnamefont {Asensio}}, \bibinfo {author}
  {\bibfnamefont {A.}~\bibnamefont {Resta}}, \bibinfo {author} {\bibfnamefont
  {B.}~\bibnamefont {Ealet}},\ and\ \bibinfo {author} {\bibfnamefont
  {G.}~\bibnamefont {Le~Lay}},\ }\bibfield  {title} {\bibinfo {title}
  {Silicene: Compelling experimental evidence for graphenelike two-dimensional
  silicon},\ }\href {https://doi.org/10.1103/PhysRevLett.108.155501} {\bibfield
   {journal} {\bibinfo  {journal} {Phys. Rev. Lett.}\ }\textbf {\bibinfo
  {volume} {108}},\ \bibinfo {pages} {155501} (\bibinfo {year}
  {2012})}\BibitemShut {NoStop}%
\bibitem [{\citenamefont {Sheverdyaeva}\ \emph {et~al.}(2017)\citenamefont
  {Sheverdyaeva}, \citenamefont {Mahatha}, \citenamefont {Moras}, \citenamefont
  {Petaccia}, \citenamefont {Fratesi}, \citenamefont {Onida},\ and\
  \citenamefont {Carbone}}]{Silicene_hybridized_NotDirac_Sanjoy2017}%
  \BibitemOpen
  \bibfield  {author} {\bibinfo {author} {\bibfnamefont {P.~M.}\ \bibnamefont
  {Sheverdyaeva}}, \bibinfo {author} {\bibfnamefont {S.~K.}\ \bibnamefont
  {Mahatha}}, \bibinfo {author} {\bibfnamefont {P.}~\bibnamefont {Moras}},
  \bibinfo {author} {\bibfnamefont {L.}~\bibnamefont {Petaccia}}, \bibinfo
  {author} {\bibfnamefont {G.}~\bibnamefont {Fratesi}}, \bibinfo {author}
  {\bibfnamefont {G.}~\bibnamefont {Onida}},\ and\ \bibinfo {author}
  {\bibfnamefont {C.}~\bibnamefont {Carbone}},\ }\bibfield  {title} {\bibinfo
  {title} {Electronic states of silicene allotropes on ag(111)},\ }\href
  {https://doi.org/10.1021/acsnano.6b07593} {\bibfield  {journal} {\bibinfo
  {journal} {ACS Nano}\ }\textbf {\bibinfo {volume} {11}},\ \bibinfo {pages}
  {975} (\bibinfo {year} {2017})},\ \Eprint
  {https://arxiv.org/abs/https://doi.org/10.1021/acsnano.6b07593}
  {https://doi.org/10.1021/acsnano.6b07593} \BibitemShut {NoStop}%
\bibitem [{\citenamefont {D{\'{a}}vila}\ \emph {et~al.}(2014)\citenamefont
  {D{\'{a}}vila}, \citenamefont {Xian}, \citenamefont {Cahangirov},
  \citenamefont {Rubio},\ and\ \citenamefont {Lay}}]{Germanene_LeLay_2014}%
  \BibitemOpen
  \bibfield  {author} {\bibinfo {author} {\bibfnamefont {M.~E.}\ \bibnamefont
  {D{\'{a}}vila}}, \bibinfo {author} {\bibfnamefont {L.}~\bibnamefont {Xian}},
  \bibinfo {author} {\bibfnamefont {S.}~\bibnamefont {Cahangirov}}, \bibinfo
  {author} {\bibfnamefont {A.}~\bibnamefont {Rubio}},\ and\ \bibinfo {author}
  {\bibfnamefont {G.~L.}\ \bibnamefont {Lay}},\ }\bibfield  {title} {\bibinfo
  {title} {Germanene: a novel two-dimensional germanium allotrope akin to
  graphene and silicene},\ }\href
  {https://doi.org/10.1088/1367-2630/16/9/095002} {\bibfield  {journal}
  {\bibinfo  {journal} {New Journal of Physics}\ }\textbf {\bibinfo {volume}
  {16}},\ \bibinfo {pages} {095002} (\bibinfo {year} {2014})}\BibitemShut
  {NoStop}%
\bibitem [{\citenamefont {Kesper}\ \emph {et~al.}(2022)\citenamefont {Kesper},
  \citenamefont {Hochhaus}, \citenamefont {Schmitz}, \citenamefont {Schulte},
  \citenamefont {Berges},\ and\ \citenamefont
  {Westphal}}]{Germanene_Ag111_2022}%
  \BibitemOpen
  \bibfield  {author} {\bibinfo {author} {\bibfnamefont {L.}~\bibnamefont
  {Kesper}}, \bibinfo {author} {\bibfnamefont {J.~A.}\ \bibnamefont
  {Hochhaus}}, \bibinfo {author} {\bibfnamefont {M.}~\bibnamefont {Schmitz}},
  \bibinfo {author} {\bibfnamefont {M.~G.~H.}\ \bibnamefont {Schulte}},
  \bibinfo {author} {\bibfnamefont {U.}~\bibnamefont {Berges}},\ and\ \bibinfo
  {author} {\bibfnamefont {C.}~\bibnamefont {Westphal}},\ }\bibfield  {title}
  {\bibinfo {title} {Tracing the structural evolution of quasi-freestanding
  germanene on {Ag}(111)},\ }\href {https://doi.org/10.1038/s41598-022-10943-0}
  {\bibfield  {journal} {\bibinfo  {journal} {Scientific Reports}\ }\textbf
  {\bibinfo {volume} {12}},\ \bibinfo {pages} {7559} (\bibinfo {year}
  {2022})}\BibitemShut {NoStop}%
\bibitem [{\citenamefont {Reis}\ \emph {et~al.}(2017)\citenamefont {Reis},
  \citenamefont {Li}, \citenamefont {Dudy}, \citenamefont {Bauernfeind},
  \citenamefont {Glass}, \citenamefont {Hanke}, \citenamefont {Thomale},
  \citenamefont {Schäfer},\ and\ \citenamefont
  {Claessen}}]{Bismuthene_SiC_2017}%
  \BibitemOpen
  \bibfield  {author} {\bibinfo {author} {\bibfnamefont {F.}~\bibnamefont
  {Reis}}, \bibinfo {author} {\bibfnamefont {G.}~\bibnamefont {Li}}, \bibinfo
  {author} {\bibfnamefont {L.}~\bibnamefont {Dudy}}, \bibinfo {author}
  {\bibfnamefont {M.}~\bibnamefont {Bauernfeind}}, \bibinfo {author}
  {\bibfnamefont {S.}~\bibnamefont {Glass}}, \bibinfo {author} {\bibfnamefont
  {W.}~\bibnamefont {Hanke}}, \bibinfo {author} {\bibfnamefont
  {R.}~\bibnamefont {Thomale}}, \bibinfo {author} {\bibfnamefont
  {J.}~\bibnamefont {Schäfer}},\ and\ \bibinfo {author} {\bibfnamefont
  {R.}~\bibnamefont {Claessen}},\ }\bibfield  {title} {\bibinfo {title}
  {Bismuthene on a {SiC} substrate: A candidate for a high-temperature quantum
  spin hall material},\ }\href {https://doi.org/10.1126/science.aai8142}
  {\bibfield  {journal} {\bibinfo  {journal} {Science}\ }\textbf {\bibinfo
  {volume} {357}},\ \bibinfo {pages} {287} (\bibinfo {year} {2017})},\ \Eprint
  {https://arxiv.org/abs/https://www.science.org/doi/pdf/10.1126/science.aai8142}
  {https://www.science.org/doi/pdf/10.1126/science.aai8142} \BibitemShut
  {NoStop}%
\bibitem [{\citenamefont {Maklar}\ \emph {et~al.}(2022)\citenamefont {Maklar},
  \citenamefont {Stühler}, \citenamefont {Dendzik}, \citenamefont {Pincelli},
  \citenamefont {Dong}, \citenamefont {Beaulieu}, \citenamefont {Neef},
  \citenamefont {Li}, \citenamefont {Wolf}, \citenamefont {Ernstorfer},
  \citenamefont {Claessen},\ and\ \citenamefont
  {Rettig}}]{Bismuthene_TRARPES_Mac22}%
  \BibitemOpen
  \bibfield  {author} {\bibinfo {author} {\bibfnamefont {J.}~\bibnamefont
  {Maklar}}, \bibinfo {author} {\bibfnamefont {R.}~\bibnamefont {Stühler}},
  \bibinfo {author} {\bibfnamefont {M.}~\bibnamefont {Dendzik}}, \bibinfo
  {author} {\bibfnamefont {T.}~\bibnamefont {Pincelli}}, \bibinfo {author}
  {\bibfnamefont {S.}~\bibnamefont {Dong}}, \bibinfo {author} {\bibfnamefont
  {S.}~\bibnamefont {Beaulieu}}, \bibinfo {author} {\bibfnamefont
  {A.}~\bibnamefont {Neef}}, \bibinfo {author} {\bibfnamefont {G.}~\bibnamefont
  {Li}}, \bibinfo {author} {\bibfnamefont {M.}~\bibnamefont {Wolf}}, \bibinfo
  {author} {\bibfnamefont {R.}~\bibnamefont {Ernstorfer}}, \bibinfo {author}
  {\bibfnamefont {R.}~\bibnamefont {Claessen}},\ and\ \bibinfo {author}
  {\bibfnamefont {L.}~\bibnamefont {Rettig}},\ }\bibfield  {title} {\bibinfo
  {title} {Ultrafast momentum-resolved hot electron dynamics in the
  two-dimensional topological insulator bismuthene},\ }\href
  {https://doi.org/10.1021/acs.nanolett.2c01462} {\bibfield  {journal}
  {\bibinfo  {journal} {Nano Letters}\ }\textbf {\bibinfo {volume} {22}},\
  \bibinfo {pages} {5420} (\bibinfo {year} {2022})},\ \Eprint
  {https://arxiv.org/abs/https://doi.org/10.1021/acs.nanolett.2c01462}
  {https://doi.org/10.1021/acs.nanolett.2c01462} \BibitemShut {NoStop}%
\bibitem [{\citenamefont {Woodruff}(2022)}]{SurfaceAlloys_Woodruf}%
  \BibitemOpen
  \bibinfo {editor} {\bibfnamefont {D.}~\bibnamefont {Woodruff}},\ ed.,\
  \href@noop {} {\emph {\bibinfo {title} {Surface Alloys and Alloy
  Surfaces}}},\ \bibinfo {edition} {1st}\ ed.,\ The Chemical Physics of Solid
  Surfaces\ (\bibinfo  {publisher} {Elsevier Science 2002},\ \bibinfo {year}
  {2022})\BibitemShut {NoStop}%
\bibitem [{\citenamefont {Vasiliev}(1997)}]{SurfAlloys_Magnetic_Vasiliev_1997}%
  \BibitemOpen
  \bibfield  {author} {\bibinfo {author} {\bibfnamefont {M.}~\bibnamefont
  {Vasiliev}},\ }\bibfield  {title} {\bibinfo {title} {Surface effects of
  ordering in binary alloys},\ }\href
  {https://doi.org/10.1088/0022-3727/30/22/002} {\bibfield  {journal} {\bibinfo
   {journal} {Journal of Physics D: Applied Physics}\ }\textbf {\bibinfo
  {volume} {30}},\ \bibinfo {pages} {3037} (\bibinfo {year}
  {1997})}\BibitemShut {NoStop}%
\bibitem [{\citenamefont {Mehendale}\ \emph {et~al.}(2010)\citenamefont
  {Mehendale}, \citenamefont {Girard}, \citenamefont {Repain}, \citenamefont
  {Chacon}, \citenamefont {Lagoute}, \citenamefont {Rousset}, \citenamefont
  {Marathe},\ and\ \citenamefont {Narasimhan}}]{AlloyFe_Au_onRu}%
  \BibitemOpen
  \bibfield  {author} {\bibinfo {author} {\bibfnamefont {S.}~\bibnamefont
  {Mehendale}}, \bibinfo {author} {\bibfnamefont {Y.}~\bibnamefont {Girard}},
  \bibinfo {author} {\bibfnamefont {V.}~\bibnamefont {Repain}}, \bibinfo
  {author} {\bibfnamefont {C.}~\bibnamefont {Chacon}}, \bibinfo {author}
  {\bibfnamefont {J.}~\bibnamefont {Lagoute}}, \bibinfo {author} {\bibfnamefont
  {S.}~\bibnamefont {Rousset}}, \bibinfo {author} {\bibfnamefont
  {M.}~\bibnamefont {Marathe}},\ and\ \bibinfo {author} {\bibfnamefont
  {S.}~\bibnamefont {Narasimhan}},\ }\bibfield  {title} {\bibinfo {title}
  {Ordered surface alloy of bulk-immiscible components stabilized by
  magnetism},\ }\href {https://doi.org/10.1103/PhysRevLett.105.056101}
  {\bibfield  {journal} {\bibinfo  {journal} {Phys. Rev. Lett.}\ }\textbf
  {\bibinfo {volume} {105}},\ \bibinfo {pages} {056101} (\bibinfo {year}
  {2010})}\BibitemShut {NoStop}%
\bibitem [{\citenamefont {Gao}\ \emph {et~al.}(2018)\citenamefont {Gao},
  \citenamefont {Sun}, \citenamefont {Lu}, \citenamefont {Li}, \citenamefont
  {Qian}, \citenamefont {Zhang}, \citenamefont {Zhang}, \citenamefont {Qian},
  \citenamefont {Ding}, \citenamefont {Lin}, \citenamefont {Du},\ and\
  \citenamefont {Gao}}]{Alloy_CuSe_onCu111}%
  \BibitemOpen
  \bibfield  {author} {\bibinfo {author} {\bibfnamefont {L.}~\bibnamefont
  {Gao}}, \bibinfo {author} {\bibfnamefont {J.-T.}\ \bibnamefont {Sun}},
  \bibinfo {author} {\bibfnamefont {J.-C.}\ \bibnamefont {Lu}}, \bibinfo
  {author} {\bibfnamefont {H.}~\bibnamefont {Li}}, \bibinfo {author}
  {\bibfnamefont {K.}~\bibnamefont {Qian}}, \bibinfo {author} {\bibfnamefont
  {S.}~\bibnamefont {Zhang}}, \bibinfo {author} {\bibfnamefont {Y.-Y.}\
  \bibnamefont {Zhang}}, \bibinfo {author} {\bibfnamefont {T.}~\bibnamefont
  {Qian}}, \bibinfo {author} {\bibfnamefont {H.}~\bibnamefont {Ding}}, \bibinfo
  {author} {\bibfnamefont {X.}~\bibnamefont {Lin}}, \bibinfo {author}
  {\bibfnamefont {S.}~\bibnamefont {Du}},\ and\ \bibinfo {author}
  {\bibfnamefont {H.-J.}\ \bibnamefont {Gao}},\ }\bibfield  {title} {\bibinfo
  {title} {Epitaxial growth of honeycomb monolayer {C}u{S}e with {D}irac nodal
  line fermions},\ }\href
  {https://doi.org/https://doi.org/10.1002/adma.201707055} {\bibfield
  {journal} {\bibinfo  {journal} {Advanced Materials}\ }\textbf {\bibinfo
  {volume} {30}},\ \bibinfo {pages} {1707055} (\bibinfo {year} {2018})},\
  \Eprint
  {https://arxiv.org/abs/https://onlinelibrary.wiley.com/doi/pdf/10.1002/adma.201707055}
  {https://onlinelibrary.wiley.com/doi/pdf/10.1002/adma.201707055} \BibitemShut
  {NoStop}%
\bibitem [{\citenamefont {Ast}\ \emph {et~al.}(2007)\citenamefont {Ast},
  \citenamefont {Henk}, \citenamefont {Ernst}, \citenamefont {Moreschini},
  \citenamefont {Falub}, \citenamefont {Pacil\'e}, \citenamefont {Bruno},
  \citenamefont {Kern},\ and\ \citenamefont {Grioni}}]{Alloy_Ag2Bi_Ag111_Ast}%
  \BibitemOpen
  \bibfield  {author} {\bibinfo {author} {\bibfnamefont {C.~R.}\ \bibnamefont
  {Ast}}, \bibinfo {author} {\bibfnamefont {J.}~\bibnamefont {Henk}}, \bibinfo
  {author} {\bibfnamefont {A.}~\bibnamefont {Ernst}}, \bibinfo {author}
  {\bibfnamefont {L.}~\bibnamefont {Moreschini}}, \bibinfo {author}
  {\bibfnamefont {M.~C.}\ \bibnamefont {Falub}}, \bibinfo {author}
  {\bibfnamefont {D.}~\bibnamefont {Pacil\'e}}, \bibinfo {author}
  {\bibfnamefont {P.}~\bibnamefont {Bruno}}, \bibinfo {author} {\bibfnamefont
  {K.}~\bibnamefont {Kern}},\ and\ \bibinfo {author} {\bibfnamefont
  {M.}~\bibnamefont {Grioni}},\ }\bibfield  {title} {\bibinfo {title} {Giant
  spin splitting through surface alloying},\ }\href
  {https://doi.org/10.1103/PhysRevLett.98.186807} {\bibfield  {journal}
  {\bibinfo  {journal} {Phys. Rev. Lett.}\ }\textbf {\bibinfo {volume} {98}},\
  \bibinfo {pages} {186807} (\bibinfo {year} {2007})}\BibitemShut {NoStop}%
\bibitem [{\citenamefont {Guan}\ \emph {et~al.}(2011)\citenamefont {Guan},
  \citenamefont {Bianchi}, \citenamefont {Bao}, \citenamefont {Perkins},
  \citenamefont {Meier}, \citenamefont {Dil}, \citenamefont {Osterwalder},\
  and\ \citenamefont {Hofmann}}]{Alloy_Bi_Ag_Marco}%
  \BibitemOpen
  \bibfield  {author} {\bibinfo {author} {\bibfnamefont {D.}~\bibnamefont
  {Guan}}, \bibinfo {author} {\bibfnamefont {M.}~\bibnamefont {Bianchi}},
  \bibinfo {author} {\bibfnamefont {S.}~\bibnamefont {Bao}}, \bibinfo {author}
  {\bibfnamefont {E.}~\bibnamefont {Perkins}}, \bibinfo {author} {\bibfnamefont
  {F.}~\bibnamefont {Meier}}, \bibinfo {author} {\bibfnamefont {J.~H.}\
  \bibnamefont {Dil}}, \bibinfo {author} {\bibfnamefont {J.}~\bibnamefont
  {Osterwalder}},\ and\ \bibinfo {author} {\bibfnamefont {P.}~\bibnamefont
  {Hofmann}},\ }\bibfield  {title} {\bibinfo {title} {Strongly enhanced
  electron-phonon coupling in the rashba-split state of the bi/ag(111) surface
  alloy},\ }\href {https://doi.org/10.1103/PhysRevB.83.155451} {\bibfield
  {journal} {\bibinfo  {journal} {Phys. Rev. B}\ }\textbf {\bibinfo {volume}
  {83}},\ \bibinfo {pages} {155451} (\bibinfo {year} {2011})}\BibitemShut
  {NoStop}%
\bibitem [{\citenamefont {Shah}\ \emph {et~al.}(2020)\citenamefont {Shah},
  \citenamefont {Sohail}, \citenamefont {Uhrberg},\ and\ \citenamefont
  {Wang}}]{AgTE_Alloy}%
  \BibitemOpen
  \bibfield  {author} {\bibinfo {author} {\bibfnamefont {J.}~\bibnamefont
  {Shah}}, \bibinfo {author} {\bibfnamefont {H.~M.}\ \bibnamefont {Sohail}},
  \bibinfo {author} {\bibfnamefont {R.~I.~G.}\ \bibnamefont {Uhrberg}},\ and\
  \bibinfo {author} {\bibfnamefont {W.}~\bibnamefont {Wang}},\ }\bibfield
  {title} {\bibinfo {title} {Two-dimensional binary honeycomb layer formed by
  {A}g and {T}e on {A}g(111)},\ }\href
  {https://doi.org/10.1021/acs.jpclett.0c00123} {\bibfield  {journal} {\bibinfo
   {journal} {The Journal of Physical Chemistry Letters}\ }\textbf {\bibinfo
  {volume} {11}},\ \bibinfo {pages} {1609} (\bibinfo {year} {2020})},\ \Eprint
  {https://arxiv.org/abs/https://doi.org/10.1021/acs.jpclett.0c00123}
  {https://doi.org/10.1021/acs.jpclett.0c00123} \BibitemShut {NoStop}%
\bibitem [{\citenamefont {Liu}\ \emph {et~al.}(2019)\citenamefont {Liu},
  \citenamefont {Liu}, \citenamefont {Miao}, \citenamefont {Xue}, \citenamefont
  {Zhang}, \citenamefont {Liu}, \citenamefont {Huang}, \citenamefont {Zhu},
  \citenamefont {Meng}, \citenamefont {Guo}, \citenamefont {Liu},\ and\
  \citenamefont {Wang}}]{AgTe_AlloySTM2019}%
  \BibitemOpen
  \bibfield  {author} {\bibinfo {author} {\bibfnamefont {B.}~\bibnamefont
  {Liu}}, \bibinfo {author} {\bibfnamefont {J.}~\bibnamefont {Liu}}, \bibinfo
  {author} {\bibfnamefont {G.}~\bibnamefont {Miao}}, \bibinfo {author}
  {\bibfnamefont {S.}~\bibnamefont {Xue}}, \bibinfo {author} {\bibfnamefont
  {S.}~\bibnamefont {Zhang}}, \bibinfo {author} {\bibfnamefont
  {L.}~\bibnamefont {Liu}}, \bibinfo {author} {\bibfnamefont {X.}~\bibnamefont
  {Huang}}, \bibinfo {author} {\bibfnamefont {X.}~\bibnamefont {Zhu}}, \bibinfo
  {author} {\bibfnamefont {S.}~\bibnamefont {Meng}}, \bibinfo {author}
  {\bibfnamefont {J.}~\bibnamefont {Guo}}, \bibinfo {author} {\bibfnamefont
  {M.}~\bibnamefont {Liu}},\ and\ \bibinfo {author} {\bibfnamefont
  {W.}~\bibnamefont {Wang}},\ }\bibfield  {title} {\bibinfo {title} {Flat
  {A}g{T}e honeycomb monolayer on {A}g(111)},\ }\href
  {https://doi.org/10.1021/acs.jpclett.9b00339} {\bibfield  {journal} {\bibinfo
   {journal} {The Journal of Physical Chemistry Letters}\ }\textbf {\bibinfo
  {volume} {10}},\ \bibinfo {pages} {1866} (\bibinfo {year} {2019})},\ \Eprint
  {https://arxiv.org/abs/https://doi.org/10.1021/acs.jpclett.9b00339}
  {https://doi.org/10.1021/acs.jpclett.9b00339} \BibitemShut {NoStop}%
\bibitem [{\citenamefont {\"Unzelmann}\ \emph {et~al.}(2020)\citenamefont
  {\"Unzelmann}, \citenamefont {Bentmann}, \citenamefont {Eck}, \citenamefont
  {Ki\ss{}linger}, \citenamefont {Geldiyev}, \citenamefont {Rieger},
  \citenamefont {Moser}, \citenamefont {Vidal}, \citenamefont {Ki\ss{}ner},
  \citenamefont {Hammer}, \citenamefont {Schneider}, \citenamefont {Fauster},
  \citenamefont {Sangiovanni}, \citenamefont {Di~Sante},\ and\ \citenamefont
  {Reinert}}]{AgTe_allyoRashbaSplit2020}%
  \BibitemOpen
  \bibfield  {author} {\bibinfo {author} {\bibfnamefont {M.}~\bibnamefont
  {\"Unzelmann}}, \bibinfo {author} {\bibfnamefont {H.}~\bibnamefont
  {Bentmann}}, \bibinfo {author} {\bibfnamefont {P.}~\bibnamefont {Eck}},
  \bibinfo {author} {\bibfnamefont {T.}~\bibnamefont {Ki\ss{}linger}}, \bibinfo
  {author} {\bibfnamefont {B.}~\bibnamefont {Geldiyev}}, \bibinfo {author}
  {\bibfnamefont {J.}~\bibnamefont {Rieger}}, \bibinfo {author} {\bibfnamefont
  {S.}~\bibnamefont {Moser}}, \bibinfo {author} {\bibfnamefont {R.~C.}\
  \bibnamefont {Vidal}}, \bibinfo {author} {\bibfnamefont {K.}~\bibnamefont
  {Ki\ss{}ner}}, \bibinfo {author} {\bibfnamefont {L.}~\bibnamefont {Hammer}},
  \bibinfo {author} {\bibfnamefont {M.~A.}\ \bibnamefont {Schneider}}, \bibinfo
  {author} {\bibfnamefont {T.}~\bibnamefont {Fauster}}, \bibinfo {author}
  {\bibfnamefont {G.}~\bibnamefont {Sangiovanni}}, \bibinfo {author}
  {\bibfnamefont {D.}~\bibnamefont {Di~Sante}},\ and\ \bibinfo {author}
  {\bibfnamefont {F.}~\bibnamefont {Reinert}},\ }\bibfield  {title} {\bibinfo
  {title} {Orbital-driven rashba effect in a binary honeycomb monolayer agte},\
  }\href {https://doi.org/10.1103/PhysRevLett.124.176401} {\bibfield  {journal}
  {\bibinfo  {journal} {Phys. Rev. Lett.}\ }\textbf {\bibinfo {volume} {124}},\
  \bibinfo {pages} {176401} (\bibinfo {year} {2020})}\BibitemShut {NoStop}%
\bibitem [{\citenamefont {Can}\ \emph {et~al.}(2021)\citenamefont {Can},
  \citenamefont {Tummuru}, \citenamefont {Day}, \citenamefont {Elfimov},
  \citenamefont {Damascelli},\ and\ \citenamefont {Franz}}]{Can21}%
  \BibitemOpen
  \bibfield  {author} {\bibinfo {author} {\bibfnamefont {O.}~\bibnamefont
  {Can}}, \bibinfo {author} {\bibfnamefont {T.}~\bibnamefont {Tummuru}},
  \bibinfo {author} {\bibfnamefont {R.~P.}\ \bibnamefont {Day}}, \bibinfo
  {author} {\bibfnamefont {I.}~\bibnamefont {Elfimov}}, \bibinfo {author}
  {\bibfnamefont {A.}~\bibnamefont {Damascelli}},\ and\ \bibinfo {author}
  {\bibfnamefont {M.}~\bibnamefont {Franz}},\ }\bibfield  {title} {\bibinfo
  {title} {High-temperature topological superconductivity in twisted
  double-layer copper oxides},\ }\href
  {https://doi.org/10.1038/s41567-020-01142-7} {\bibfield  {journal} {\bibinfo
  {journal} {Nature Physics}\ }\textbf {\bibinfo {volume} {17}},\ \bibinfo
  {pages} {519} (\bibinfo {year} {2021})}\BibitemShut {NoStop}%
\bibitem [{\citenamefont {Jiang}\ \emph {et~al.}(2021)\citenamefont {Jiang},
  \citenamefont {Liu}, \citenamefont {Xing}, \citenamefont {Liu}, \citenamefont
  {Guo}, \citenamefont {Liu},\ and\ \citenamefont {Zhao}}]{Jiang21}%
  \BibitemOpen
  \bibfield  {author} {\bibinfo {author} {\bibfnamefont {X.}~\bibnamefont
  {Jiang}}, \bibinfo {author} {\bibfnamefont {Q.}~\bibnamefont {Liu}}, \bibinfo
  {author} {\bibfnamefont {J.}~\bibnamefont {Xing}}, \bibinfo {author}
  {\bibfnamefont {N.}~\bibnamefont {Liu}}, \bibinfo {author} {\bibfnamefont
  {Y.}~\bibnamefont {Guo}}, \bibinfo {author} {\bibfnamefont {Z.}~\bibnamefont
  {Liu}},\ and\ \bibinfo {author} {\bibfnamefont {J.}~\bibnamefont {Zhao}},\
  }\bibfield  {title} {\bibinfo {title} {Recent progress on 2d magnets:
  Fundamental mechanism, structural design and modification},\ }\href
  {https://doi.org/10.1063/5.0039979} {\bibfield  {journal} {\bibinfo
  {journal} {Applied Physics Reviews}\ }\textbf {\bibinfo {volume} {8}},\
  \bibinfo {pages} {031305} (\bibinfo {year} {2021})}\BibitemShut {NoStop}%
\bibitem [{\citenamefont {Ne\v{c}as}\ and\ \citenamefont
  {Klapetek}(2012)}]{Gwyddion}%
  \BibitemOpen
  \bibfield  {author} {\bibinfo {author} {\bibfnamefont {D.}~\bibnamefont
  {Ne\v{c}as}}\ and\ \bibinfo {author} {\bibfnamefont {P.}~\bibnamefont
  {Klapetek}},\ }\bibfield  {title} {\bibinfo {title} {Gwyddion: an open-source
  software for {SPM} data analysis},\ }\href
  {https://doi.org/10.2478/s11534-011-0096-2} {\bibfield  {journal} {\bibinfo
  {journal} {Central European Journal of Physics}\ }\textbf {\bibinfo {volume}
  {10}},\ \bibinfo {pages} {181} (\bibinfo {year} {2012})}\BibitemShut
  {NoStop}%
\bibitem [{\citenamefont {Guo}\ \emph {et~al.}(2022)\citenamefont {Guo},
  \citenamefont {Dendzik}, \citenamefont {Grubi\v{s}i\'{c}-\v{C}abo},
  \citenamefont {Berntsen}, \citenamefont {Li}, \citenamefont {Chen},
  \citenamefont {Matta}, \citenamefont {Starke}, \citenamefont {Hessmo},
  \citenamefont {Weissenrieder},\ and\ \citenamefont {Tjernberg}}]{BALTAZAR}%
  \BibitemOpen
  \bibfield  {author} {\bibinfo {author} {\bibfnamefont {Q.}~\bibnamefont
  {Guo}}, \bibinfo {author} {\bibfnamefont {M.}~\bibnamefont {Dendzik}},
  \bibinfo {author} {\bibfnamefont {A.}~\bibnamefont
  {Grubi\v{s}i\'{c}-\v{C}abo}}, \bibinfo {author} {\bibfnamefont {M.~H.}\
  \bibnamefont {Berntsen}}, \bibinfo {author} {\bibfnamefont {C.}~\bibnamefont
  {Li}}, \bibinfo {author} {\bibfnamefont {W.}~\bibnamefont {Chen}}, \bibinfo
  {author} {\bibfnamefont {B.}~\bibnamefont {Matta}}, \bibinfo {author}
  {\bibfnamefont {U.}~\bibnamefont {Starke}}, \bibinfo {author} {\bibfnamefont
  {B.}~\bibnamefont {Hessmo}}, \bibinfo {author} {\bibfnamefont
  {J.}~\bibnamefont {Weissenrieder}},\ and\ \bibinfo {author} {\bibfnamefont
  {O.}~\bibnamefont {Tjernberg}},\ }\bibfield  {title} {\bibinfo {title} {A
  narrow bandwidth extreme ultra-violet light source for time- and
  angle-resolved photoemission spectroscopy},\ }\href
  {https://doi.org/10.1063/4.0000149} {\bibfield  {journal} {\bibinfo
  {journal} {Structural Dynamics}\ }\textbf {\bibinfo {volume} {9}},\ \bibinfo
  {pages} {024304} (\bibinfo {year} {2022})},\ \Eprint
  {https://arxiv.org/abs/https://doi.org/10.1063/4.0000149}
  {https://doi.org/10.1063/4.0000149} \BibitemShut {NoStop}%
\bibitem [{\citenamefont {Hoffmann}\ \emph {et~al.}(2002)\citenamefont
  {Hoffmann}, \citenamefont {Lunt}, \citenamefont {Jones}, \citenamefont
  {Field},\ and\ \citenamefont {Ziesel}}]{SGM3beamline}%
  \BibitemOpen
  \bibfield  {author} {\bibinfo {author} {\bibfnamefont {S.~V.}\ \bibnamefont
  {Hoffmann}}, \bibinfo {author} {\bibfnamefont {S.~L.}\ \bibnamefont {Lunt}},
  \bibinfo {author} {\bibfnamefont {N.~C.}\ \bibnamefont {Jones}}, \bibinfo
  {author} {\bibfnamefont {D.}~\bibnamefont {Field}},\ and\ \bibinfo {author}
  {\bibfnamefont {J.-P.}\ \bibnamefont {Ziesel}},\ }\bibfield  {title}
  {\bibinfo {title} {An undulator-based spherical grating monochromator
  beamline for low energy electron-molecule scattering experiments},\ }\href
  {https://doi.org/10.1063/1.1517143} {\bibfield  {journal} {\bibinfo
  {journal} {Review of Scientific Instruments}\ }\textbf {\bibinfo {volume}
  {73}},\ \bibinfo {pages} {4157} (\bibinfo {year} {2002})},\ \Eprint
  {https://arxiv.org/abs/https://doi.org/10.1063/1.1517143}
  {https://doi.org/10.1063/1.1517143} \BibitemShut {NoStop}%
\end{thebibliography}%
	
\end{document}